\documentclass[useAMS,usenatbib]{mn2e}
\usepackage{textcomp,graphicx,amsmath,pdflscape,amssymb,subfigure}
\usepackage{epsfig}
\usepackage{amsmath}

\usepackage{epsf}
\usepackage{lscape}
\usepackage{rotating}
\usepackage{multirow}
\usepackage{subfigure}
\newcommand\aj{{AJ}}%
\newcommand\apj{{ApJ}}%
\newcommand\apjl{{ApJ}}%
\newcommand\apjs{{ApJS}}%
\newcommand\ao{{Appl.~Opt.}}%
\newcommand\aap{{A\&A}}%
\newcommand\mnras{{MNRAS}}%
\def\mathbi#1{\textbf{\em#1}}

\begin{document}

\title[Follow-Up strategy of low-cadence photometry]{Directed follow-up strategy of low-cadence photometric surveys in Search of transiting exoplanets - I. Bayesian approach for adaptive scheduling} 
%{\large \textbf{I. Bayesian approach for adaptive scheduling}}}

\author[Y. Dzigan and S. Zucker]
  {Yifat Dzigan$^1$\thanks{E-mail: yifatdzigan@gmail.com}
  and Shay Zucker$^1$\thanks{E-mail: shayz@post.tau.ac.il}\\
$^1$Department of Geophysics and Planetary Sciences, Tel Aviv University, Tel Aviv 69978, Israel\\
 }
\date{Accepted for publication in Monthly Notices of the Royal Astronomical Society Main Journal. Accepted 2011 April 8.}
\pagerange{\pageref{firstpage}--\pageref{lastpage}} \pubyear{2011}

\maketitle
\label{firstpage}

\begin{abstract}
We propose a novel approach to utilize low-cadence photometric
surveys for exoplanetary transit search. Even if transits
are undetectable in the survey database alone, it can still be
useful for finding preferred times for directed follow-up
observations that will maximize the chances to detect transits.
We demonstrate the approach through a few simulated cases.
These simulations are based on the \textit{Hipparcos} Epoch Photometry
data base, and the transiting planets whose transits were already
detected there. In principle, the approach we propose will be
suitable for the directed follow-up of the photometry from the planned
\textit{Gaia} mission, and it can hopefully significantly increase the yield
of exoplanetary transits detected, thanks to \textit{Gaia}.
\end{abstract}


\section{Introduction}\label{intro}

The idea to detect transits of exoplanets in the \textit{Hipparcos} Epoch
Photometry \citep{1997yCat.1239....0E} trigerred several studies that checked the feasibility of
such an attempt. \cite{2006A&A...445..341H} concluded that \textit{Hipparcos}
photometry did not look like an efficient tool for transit detection,
without using any prior information. Indeed, some teams have
made \textit{posterior} detections of the transits of HD 209458
\citep[][]{2000A&A...355..295R, 2000ApJ...532L..51C,
1999IBVS.4816....1S}, and of HD 189733 \citep{2005A&A...444L..15B}, a
detection that \cite{2006A&A...445..341H} confirmed. Those teams used
the previously available knowledge of the orbital elements of the
exoplanets (especially the period and the transit phase), in order to
phase the \textit{Hipparcos} data of those stars. Using the large time span
that had elapsed since the \textit{Hipparcos} observations (for example, about
830 orbital periods between \textit{Hipparcos} observations and the
observations that \cite{2000ApJ...532L..51C} used for HD 209458), the
teams managed to drastically reduce the uncertainties of the known periods.
The posterior detections of both transits prove that information about the transits does exist in the data, although it is obviously futile to try to detect the transits in the na\"ive Box Least Squares (BLS) approach \citep{2002A&A...391..369K}. Thus the posterior detections motivated us to re-examine
\textit{Hipparcos} Epoch Photometry data and to look for a way to utilize this
survey and similar low-cadence photometric surveys, to detect
exoplanets. The approach we
propose here is to use the data to maximize the chances to detect
transits during hypothetical follow-up campaigns, i.e., instead of attempting to detect a transit, we use the data to schedule follow-up observations that together with the old data set may enable its detection.

In order to maximize the probability of sampling a transit in those
surveys, and in order to predict the best possible future observing
times, we chose to use Bayesian inference methods.

Bayesian analysis is based on Bayes theorem and can be written as: 

\begin{equation}\label{Bayes}
 p(H_i|D,I)=\frac{p(H_i|I)p(D|H_i,I)}{p(D|I)},
\end{equation}
where $p(H_i|D,I)$ is the posterior probability of the Hypothesis $H_i$, given the prior information, $I$, and the data, $D$.

$p(D|H_i,I)$ is the probability of obtaining the data $D$, given that $H_i$ and $I$ are true. It is also known as the likelihood function $L(H_i)$. 

$p(D|I)=\sum_i p(H_i|I)p(D|H_i,I)$ is a normalization factor that ensures that $\sum_i p(H_i|D,I)=1$. It is usually referred to as the prior predictive probability for $D$, or the global likelihood for the entire class of hypotheses.

In the Bayesian framework we start from a prior knowledge we introduce into the prior probability distribution, $p(H_0|I)$. The choice of prior distribution can affect the posterior distribution, especially if our observed data do not strongly constrain the model parameters. If our prior knowledge is poor, $p(H_0|I)$ can spread over a wide range of possible values for the model parameters. 

Whenever new data are available, it is possible to incorporate the new data in our model through the likelihood function, combined with the prior, to obtain a new posterior density probability, $p(H_0|D_1,I)$, for the parameter. As soon as we obtain another set of data, $D_2$, we recalculate the posterior density probability in order for it to reflect our new state of knowledge. The possibility to combine new data sets into the original data we have will allow us to accomplish our goal of detecting transiting exoplanets using scheduled follow-up observations.

\cite{2005ApJ...631.1198G}, \cite{2006ApJ...642..505F} and others have
already shown that Bayesian inference is a useful tool for analyzing
precise radial velocity (RV) data of planet-hosting stars.
\cite{2007MNRAS.381.1607G} used Bayesian inference model selection
for the problem of multiple planets, and \cite{2005ApJ...631.1198G}
used it to construct posterior probability density functions
of the light-curve parameters.
\cite{2001A&A...365..330D} and \cite{2002A&A...395..625A} demonstrated
the use of the Bayesian approach to study planetary transits.

Our implementation of Bayesian inference is based on the
Metropolis-Hastings (MH) algorithm, which is a version of the more general
Markov-Chain Monte Carlo (MCMC) approach \citep{2005blda.book.....G}.

A Markov chain is calculated using an initial set of parameter values, $\overline{X}_0$, and a transition probability, $p(\overline{X}_{n+1}|\overline{X}_n,I)$, that describes the probability of moving from the current state to the next one. The transition probability depends on the acceptance probability, described later in Section \ref{MCMC_approach}, and if properly constructed, then after excluding the so-called ``burning time'', we can use the chain as a sample from the desired distribution.
MH Algorithm is an implementation of the MCMC that is used for obtaining a sequence of random samples from a probability distribution. 

The MH algorithm does not require good initial guess of the parameters values in order to estimate the posterior distribution. This is one of the most important advantages of the algorithm. The algorithm is capable of exploring all regions of the parameter space having significant probabilities (assuming it meets several basic requirements). The analysis also yields the marginal posterior probability distribution functions for each of the model parameters, and their uncertainties.

In Section \ref{directedFW} we describe the follow-up approach we developed
to detect transiting exoplanets, based on observations from low cadence surveys.
In Section \ref{MCMC_approach} we give a brief review of Bayesian inference and its applications
for our follow-up strategy. Sections \ref{HD209458_5pars} and \ref{HD189733_5pars} demonstrate the approach
by applying it on two stars that are known to harbor hot-Jupiters, HD 209458 and HD 189733, using the \textit{Hipparcos} data base. Section \ref{sanity_check} shows some ``sanity checks`` we performed on data that do not contain the transit signal at all.
We conclude and describe future applications of the strategy in Section \ref{cuncluding}.

\section{Directed follow-up}\label{directedFW}

Our ultimate goal is to detect transiting exoplanets using follow-up
observations, which will be carefully scheduled to increase the
chances to capture transits, should they exist. Thus, our approach
does not focus on obtaining a detailed transit model that best fits
the available data, but on building a probability distribution
function of the parameters of a simple model, based on these data. The
transit model that we use in this work is a very simplistic one, based on
the BLS philosophy \citep{2002A&A...391..369K}. Thus, we model a
transit light curve as a box-shaped transit with two phases, in and
out of transit, and ignore the duration of the ingress and egress
phases, as well as the details of the limb darkening. These details
are less relevant in low-precision, low-cadence surveys, and using
fewer parameters makes the model more robust. We use the Bayesian
MH algorithm to obtain a posterior probability
distribution for the model parameters, and then use this distribution
to prioritize the timing of the observations of the chosen stars for
follow-up observations.

The directed follow-up approach is not suitable for space missions
like \textit{Corot} or \textit{Kepler}. Such missions, due to their high cadence, will
not benefit from the approach since their phase and period coverage are
already quite complete.  Instead, we aim for all-sky surveys like
\textit{Hipparcos} \citep{1997A&A...323L..61V}, or its successor, \textit{Gaia}
\citep{2006MNRAS.367..290J}, in order to use their extensive
low-cadence photometric databases for exoplanets search.

A simplified (BLS-like) transit light curve is parametrized by five quantities,
e.g., the period, phase, and width of the transit, and the
flux levels in-transit and ex-transit. The first step in our proposed
procedure is to apply the MH algorithm to the \textit{Hipparcos} measurements of a target
star. This results in five Markov chains that include the successful iterations
for each one of the parameters.  A successful iteration is one that was
accepted by the MH algorithm. After removing the ``burning time'', 
each chain represents the stationary
distribution of the parameters, which we use as their estimated
Bayesian posterior distributions, for our current state of knowledge
\citep{2005blda.book.....G}. Unlike the case of precise high-cadence
surveys, even if the star does host a transiting exoplanet, due to the
low precision and low cadence of the observations we do not expect the
distribution to concentrate around a single solution, but rather show
different periods that might fit the data, besides the unknown correct one.

The next step of our procedure is to assign each point in time the
probability that a transit will occur at that time. Calculating this
probability is easy using the posterior distributions we found in the
first stage. Basically for time $t$, we count the number of MCMC successful iterations whose values of $P$, $T_c$ and $w$ predict a transit in time $t$. Normalizing this number by the number of total iterations yield the Instantaneous Transit Probability (ITP) for time $t$. 
If the ITP has significantly high values for
certain times, then a follow-up observation is worthwhile at those
preferred times.  

When we examine the ITP function of different
observations and simulations we performed, it is clear that the shape
of this function when a transit signal exists contains sharp peaks, where the probability of transit is relatively high.
This behavior is crucial for defining preferred times for follow-up
observations. If we sample the values of the predicted probabilities
we can test for this behavior by the skewness of this sample. 
The skewness of a random variable is generally defined as 
\begin{equation}
 S=\frac{\langle (x-\langle \hat{x}\rangle)^3\rangle}{\sigma^3} \ ,
\end{equation}
Where $\langle \cdot \rangle$ denotes the operation of averaging over
the sample values.

The skewness of the ITP is actually a measure of the
amount of outliers, where the
term outliers actually refers to peaks with significantly high values.
The absence of such 'outliers' means that there are no preferred times for follow-up
observation; thus the ITP values will be more
symmetrically distributed, with a skewness value close to zero ($S=0$
for a normal distribution).

In order to prioritize the
stars for follow-up observations, we need to rank them.
We propose to use the skewness of the ITP as a
criterion for prioritizing stars for follow-up observations, together with the actual ITP values for follow-up times predictions.

Another criterion we propose for the prioritization process is to use the Wald
test for the posterior probability distribution of the transit depth,
which is a measure of the 'strength' of the signal we are looking for.
The Wald test, named after Abraham Wald, is the most simplistic
statistical test designed to examine the acceptance of a hypothesis
\citep[e.g.][]{2006sppp.conf.....L}. The Wald statistic for a random variable
$x$ is defined by the simple expression:
\begin{equation}
\label{Wald}
W=\frac{E(x) -x_0}{\mathrm{std}(x)}.  
\end{equation} where $E(x)$ and
$\mathrm{std}(x)$ are the expected value and standard deviation correspondingly of the
variable $x$, and $x_0$ is the nominal value of $x$
according to the null hypothesis. In our case $x$ is the transit depth,
and the moments ($E$ and $\mathrm{std}$) are calculated based on the
posterior probability distribution. $x_0$ is simply zero, since our
null hypothesis is the absence of any transit. In a sense, the Wald
statistic for the transit depth quantifies the degree to which we
believe there is a periodic transit-like dimming of the star, based on
the available photometry. 
A high value of the Wald statistic indicates a relatively narrow posterior distribution of the transit depth. This may indicate that there are periods according to which the low-flux measurements, corresponding to a transit-like dimming, are relatively concentrated in a short phase. This short phase can be the hypothetical transit which we look for.

Performing the follow-up observations at the times directed by the
previous step is the final step of the strategy. A combination of both
the `old' data from the survey and the new observations at the
directed time eliminates periods that do not fit our new state of
knowledge. The procedure is repeated until we detect a transiting
planet, or exclude its existence.

In Sections \ref{HD209458_5pars} and \ref{HD189733_5pars} we examine the strategy for two known transiting planets, HD 209458b and HD 189733b, using the \textit{Hipparcos} photometric catalogue. The promising results show that the strategy is efficient in utilizing low-cadence low-precision surveys for exoplanets transit search.

\section{Bayesian approach - simplified transit model}\label{MCMC_approach}

Inspired by the BLS \citep{2002A&A...391..369K}, our model is a simple box-shaped transit light-curve, 
with five parameters that characterize it: ${\mathbi{X}}=\{P,T_c,w,d,m\}$, where $P$ is the orbital period, $T_c$ is the time of mid-transit, $w$ is the transit duration, $d$ is the depth of the eclipse and $m$ is the mean magnitude out of transit.

Let $v_k$ and $\sigma_k$ denote the observed magnitude and its
associated uncertainty at time $t_k$, respectively. Let $m$ denote
the magnitude out of transit, which is assumed to be constant.

Assuming a simple 'white' Gaussian model for the observations, we can
write down the likelihood function explicitly:
\begin{equation*}
\begin{array}{ll}
p(D|\mathbi{X})=\prod_k \frac{1}{\sqrt{2\pi}\sigma_k} 
\exp \left(-\frac{(v_k-\mu_k)^2}{2\sigma_k^2} \right) \\
= \frac{1}{(2\pi)^\frac{K}{2}} \prod_k \frac{1}{\sigma_k} \exp \left(-  \sum \frac{(v_k-\mu_k)^2}{2\sigma_k^2} \right)\ ,
\end{array}
\end{equation*}
where
\begin{equation}
\mu_k = \left\{
	\begin{array}{ll}
	m & \textrm{if $t_k$ is out of transit,} \\
	m+d & \textrm{if $t_k$ is in transit,}
	\end{array} \right.
\end{equation}\label{statistics1}
and $K$ is the number of observations. (Note that the magnitude during transit is defined as $m+d$, since we use magnitude units and not flux units).
The exponent in equation (4) is actually half the well-known
$\chi^2$ statistic.

The MH algorithm can now be summarized by the following description:

\noindent
1. Initialize \mbox{\boldmath$X_0$} -- the initial guess for the set of model parameters; set $n=0$.

\noindent
2. Draw a sample $\mathbi{Y}$ (trial state) from a proposal
distribution $q(\mathbi{Y}|\mbox{\boldmath$X_0$})$. This distribution can
be a Gaussian distribution, centered around \mbox{\boldmath$X_n$} -- the
current set of model parameter values.  

The acceptance probability,
$\alpha$, is defined by: $\alpha=\min(1,r)$\label{acceptance
probability}, where
\begin{equation}
r=\frac{p \left(\mathbi{Y} \right) p \left(D|\mathbi{Y} \right)}{p(\mbox{\boldmath$X_n$})p(D|\mbox{\boldmath$X_n$})}\frac{q(\mbox{\boldmath$X_n$}|\mathbi{Y})}{q(\mathbi{Y}|\mbox{\boldmath$X_n$})},
\end{equation}\label{Metropolis-ratio}
is the Metropolis ratio, which is composed from the prior$\times$likelihood and the
proposal distributions.  If the proposal distribution is symmetric,
then the second factor in the Metropolis ratio is equal to 1.

\noindent
3. Sample a random variable,$u$, from a uniform distribution, in the interval $0-1$.

\noindent
4. If $u \leq \alpha$ set \mbox{\boldmath${X}_{n+1}$}=$\mathbi{Y}$ (a successful iteration),
	   else set \mbox{\boldmath${X}_{n+1}$}=\mbox{\boldmath$X_n$}.

This last step results in accepting the new trial state $Y$ with probability $\alpha$.

\noindent
5. $n=n+1$.

\noindent
6. Go back to step 2.

Steps $2-6$ are repeated $N-1$ times to produce a Markov chain of length $N$.

For a wide range of proposal distributions, $q(\mathbi{Y}|\mbox{\boldmath$X_n$})$, after an initial burn-in period 
(which is discarded), the algorithm creates samples of \mbox{\boldmath$X_n$} with a probability density function 
that covers the desired range, the posterior distribution, $p(\mbox{\boldmath$X_n$}|D)$.

\subsection{Choice of priors}

The choice of prior distributions is important in Bayesian analysis, as a non-educated choice can produce misleading results \citep{2009MNRAS.394.1936B}.

For the orbital period, we adopt the approach proposed by \cite{2006ApJ...642..505F} and use a uniform prior in $\log{P}$,

\begin{equation}
 p(P)=\frac{1}{P \ln \left( \frac{P_{max}}{P_{min}} \right)} \ .
\end{equation}
The theoretical lower limit of the orbital period according to the
Roche limit $ \left( d\approx 2.423\times R_s
\sqrt[3]{\frac{\rho_s}{\rho_p}} \right)$ for a planet with $m_p\sim 10
M_{Jup}$, orbiting a star with a solar mass, is approximately 0.2 d
\citep{2007ASPC..371..189F}, while the upper limit can be set at
around $10^3$ d, or about three times the duration of the data
\citep{2005ApJ...631.1198G}. Since the time-span of the
data is long (for example, \textit{Hipparcos} data spans more than a 1000 d), we
choose an upper limit of the order of the time-span of the data, which
is much longer than the period of any known transiting exoplanet.

We use the same form of prior for the transit duration
\begin{equation}
 p(w)=\frac{1}{w \ln \left( \frac{w_{max}}{w_{min}} \right)},
\end{equation}
where we chose, somewhat arbitrarily, a lower limit of $w=0.001$ d
and an upper limit of 1 d. This range includes all known
exoplanetary transit durations.  The other three parameters of the
transit model $(T_c, d, m)$ were assigned a uniform prior, where $T_c$
and $m$ have the range of the data as their upper and lower limits,
and the transit depth, $d$, is uniformly distributed between $0$ and
$1$.  The prior distribution in our problem, assumping that
the parameters are independent, can be described by 
\begin{equation}
 p(\mbox{\boldmath$X_n$})=p(P)p(T_c)p(w)p(d)p(m).
\end{equation}
The most significant dependence expected between the orbital parameters is between $P$ and $w$, since the maximum value of $w$ can be related to $P$. The strong dependence of $w$ on the orbital inclination 'masks out' this correlation which is why we chose to ignore it in this work. In future applications we may use more complicated priors, which might include dependence among the parameters.

\subsection{The Proposal Distribution}

We choose the proposal distribution for each parameter individually.
For $T_c$, $d$ and $m$ we use Gaussian proposal distributions centered
around the last value in the Markov chain.  For the transit duration,
$w$, which has to be between $0$ and around a few hours, we choose a
lognormal distribution for the proposal distribution in order
to avoid negative values.

The period, $P$, requires special considerations.  The structure of
many kinds of periodograms shows that the likelihood function, when
seen as a function of the period, has a very complex structure of
sharp local maxima and minima. Even for good-quality data, we expect
the likelihood to have sharp peaks in harmonics and subharmonics of
the correct period. We propose to use this shortcoming to our
advantage, by using a ``jumping'' proposal distribution for $\log
P$. Thus, besides the small steps around the previous value of the Markov chain, we
propose to allow, in some specified probabilities, jumps to a period
which is a multiple or a divisor of the current period. The probability to move to an integer multiple or divisor of the current value of the period is $Prob=1/10$, while in random probability of $9/10$, the moves are the usual ones around the current value of the Markov chain. This should allow a more efficient exploration of the parameter space.

\section{HD 209458}\label{HD209458_5pars}

We first apply the strategy on the \textit{Hipparcos} Epoch Photometry of HD
209458. \textit{Hipparcos} observed HD 209458 (HIP 108859) in non-uniformly
distributed $89$ epochs, in a time-span of about $1084$ d. We use
all the data points since they all have a quality flag $\le2$, which
means they were accepted by at least one of the two data reduction
consortia \citep{1997ESASP1200.....P}. The estimated standard errors
of the individual $H_p$ magnitudes are around $0.015$ mag, which is of
the order of the transit depth (the signal we are looking for).

\begin{figure}
%\begin{center}
\subfigure{
\includegraphics[width=0.5\textwidth]{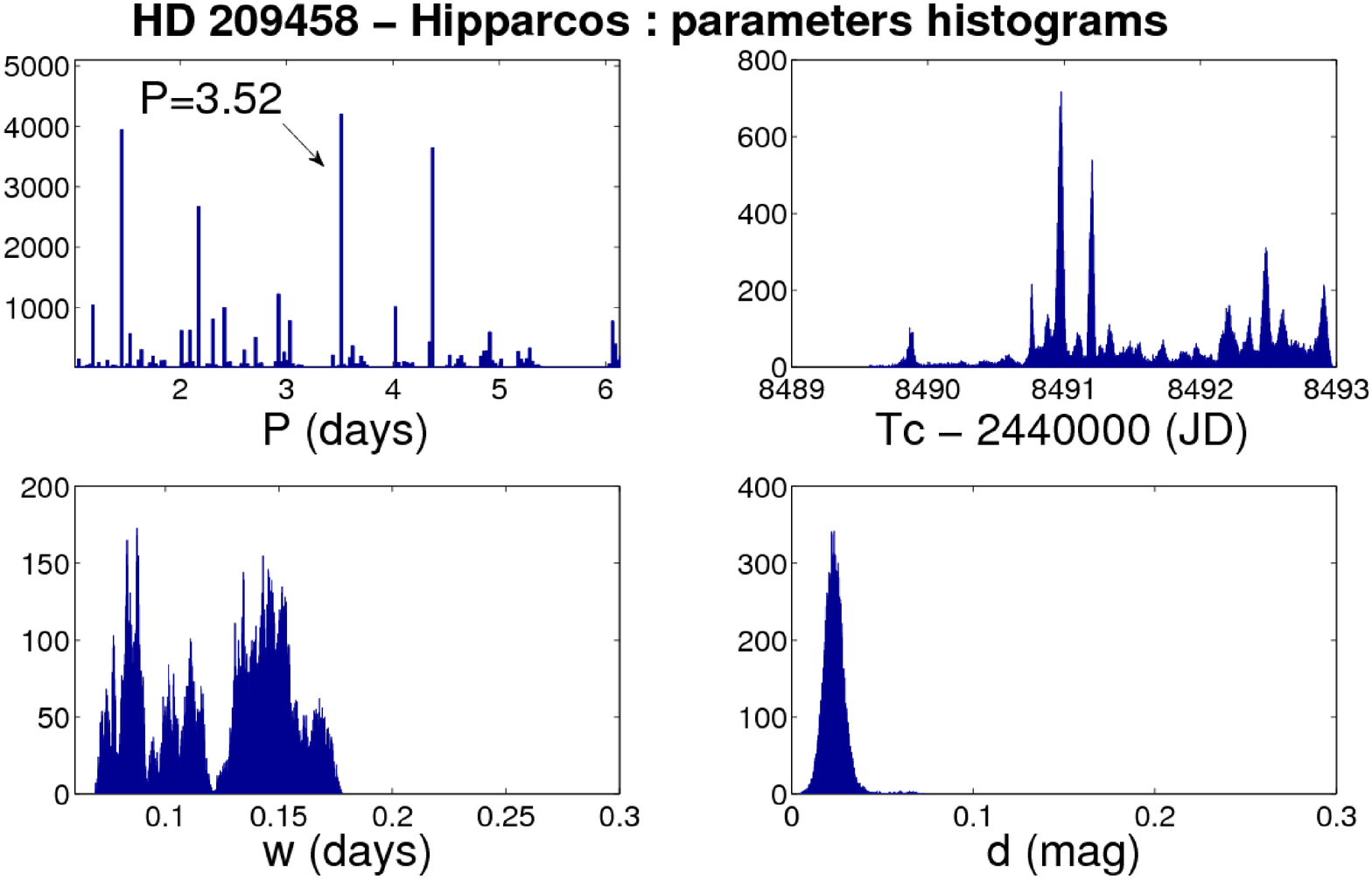}}
\subfigure{
\includegraphics[width=0.5\textwidth]{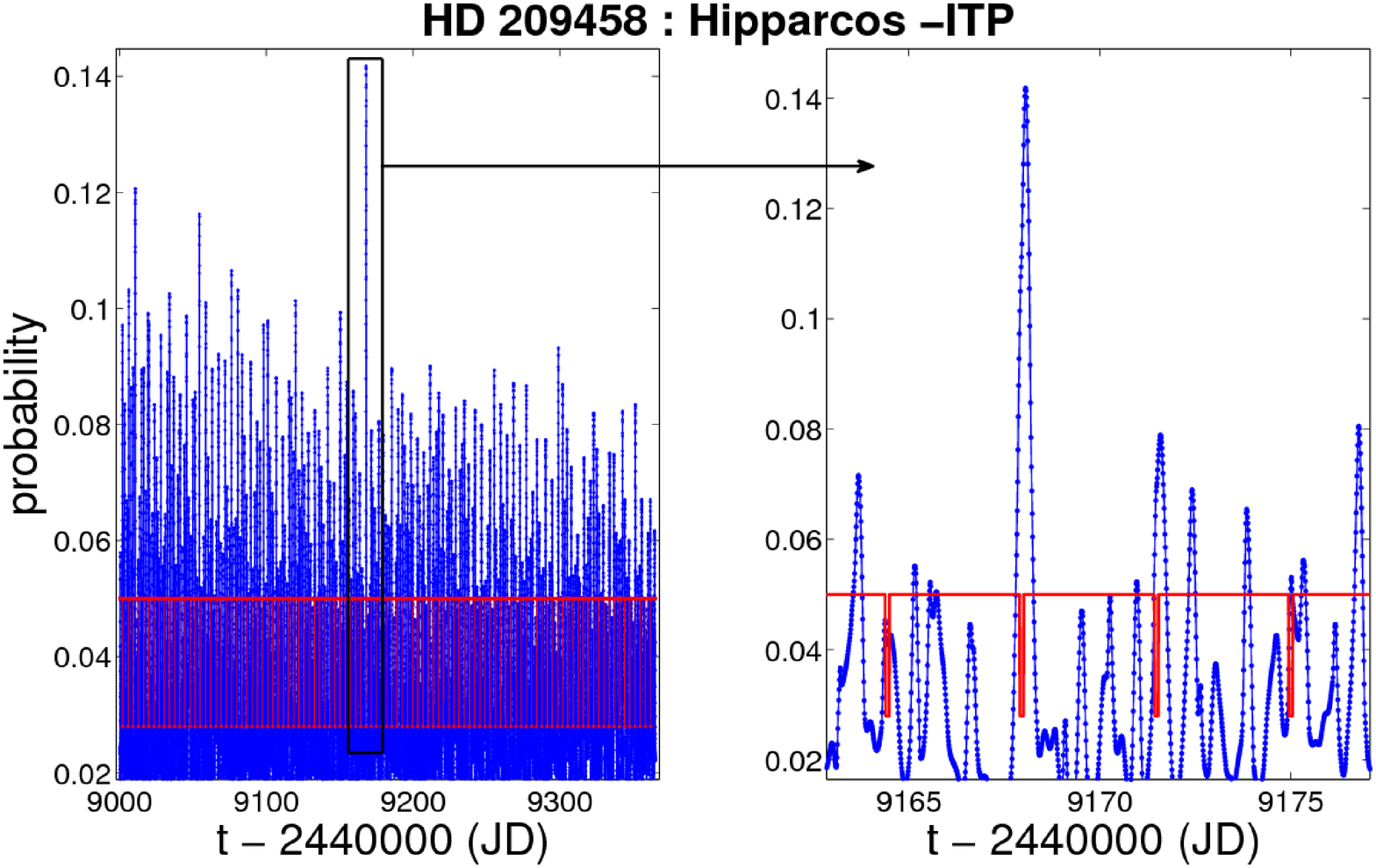}}
\caption{Top: HD 209458 - Histograms of the posterior probability distribution functions of four orbital parameters found using the MH algorithm for the \textit{Hipparcos} measurements of the star: the period $P$, time of mid-transit $Tc$, transit duration $w$ and the depth of the transit $d$. Bottom: ITP function for the first year after \textit{Hipparcos} observations, compared with the known transit light curve (orbital parameters derived using \citet{2000ApJ...532L..51C}). The significant peaks of the ITP fit mid-transit time; therefore, a single follow-up observation could have detected the transit.}
\label{fig.HD209458hist}
%\end{center}
\end{figure}

In the Bayesian framework we choose priors for the parameters as
described in Section \ref{MCMC_approach}, and allow the algorithm to explore
the parameter space in order to find the different solutions that fit
the data. The resulting posterior distributions for the relevant
parameters are shown in Fig.~\ref{fig.HD209458hist}. As can be seen
from the period histogram, the distribution does not favor any single period,
but rather has several distinctive peaks.
The most likely period is $P\approx 3.52$ d, which is consistent
with the known period of HD 209458 \citep{2000ApJ...532L..51C}, while the other probable periods
are different periods that fit the data as well.

We performed the Wald test to test the hypothesis of the presence of a
planet that is transiting the star and found that the expected value
of the transit depth posterior distribution in our analysis is
$E(d) = 0.02$ mag, and the value of the Wald test (equation \ref{Wald}) is
$W=4.22$, a result with a 4 $\sigma$ significance.

We continue with the second part of the strategy- examining the most
probable time to observe the star in a follow-up observation. Our
simulated directed follow-up explores one year that began a month
after the last observation of \textit{Hipparcos} and found the best times to
observe the star in order to sample the transit. The follow-up
predictions are shown on the bottom-right panel of
Fig.~\ref{fig.HD209458hist} (the ITP function). We add the known transit light-curve (based on the orbital parameters derived by \citet{2000ApJ...532L..51C}) to
the figure for comparison between the predictions and the actual
transits. The time that was most preferred by our predictions indeed
fits inside a transit, meaning it would have been possible to detect the transit
with only one follow-up observation conducted after \textit{Hipparcos} using
our proposed strategy. We examined the skewness of the ITP for prioritization purposes and found it to be $S=1.4$.

As will be shown in Section \ref{HD189733_5pars}, and also by Table \ref{table.waldvsSkewness}, the ITP we obtained from \textit{Hipparcos} data of HD 209458
is an exception, with a relatively low skewness, while other ITP
skewness values for data that contain a transit signal
are usually higher. We can understand this anomaly by looking at the
ITP of the star in question: there are
many peaks of the follow-up probability, and they are distributed over
the entire range we examined (one year post-\textit{Hipparcos}), with
significant values for follow-up (above $0.1$). The relatively symmetric distribution of the ITP is the
cause of the low skewness value.  The significance of the ITP for the \textit{Hipparcos} data of 
HD 209458 is still high enough
for a follow-up observation to be worthwhile, and together with the
Wald statistic, the star would have gotten a high priority for
follow-up observations, despite the relatively low skewness.

\begin{figure}
%\begin{center}
\includegraphics[width=0.5\textwidth]{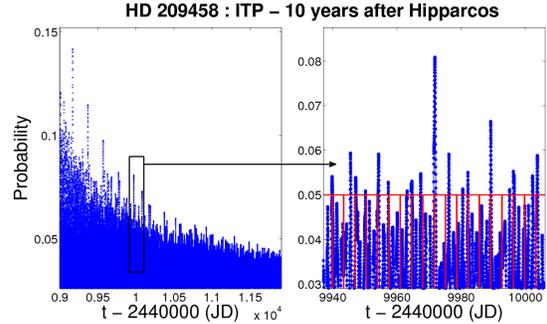}
\caption{HD 209458: ITP function for 10 yr after \textit{Hipparcos} observations. The probability of sampling a transit smears as the time elapsed from the observations, but even three years after \textit{Hipparcos}, a transit detection was possible.}
\label{fig.HD209458followup10years}
%\end{center}
\end{figure}

We also check the follow-up predictions for
$10$ yr after \textit{Hipparcos}, as shown in
Fig.~\ref{fig.HD209458followup10years}. Long after the last
observation, the ITP decreases, although
some peaks remain, and when looking carefully, we can see that even
three years after \textit{Hipparcos}, we could have detected the transit using
our follow-up predictions.

The final step of the strategy described in Section \ref{directedFW}, is to perform 
follow-up observations according to the most significant peak of the ITP.
Since the time that has elapsed since \textit{Hipparcos} cause the ITP peaks to be smeared, we cannot perform current follow-up observations
for significant ITP peaks found using the algorithm, so we simulated such observations, and then combined them
with the \textit{Hipparcos} data, to recalculate the ITP. 
\begin{figure}
%\begin{center}
\subfigure{
\includegraphics[width=0.5\textwidth]{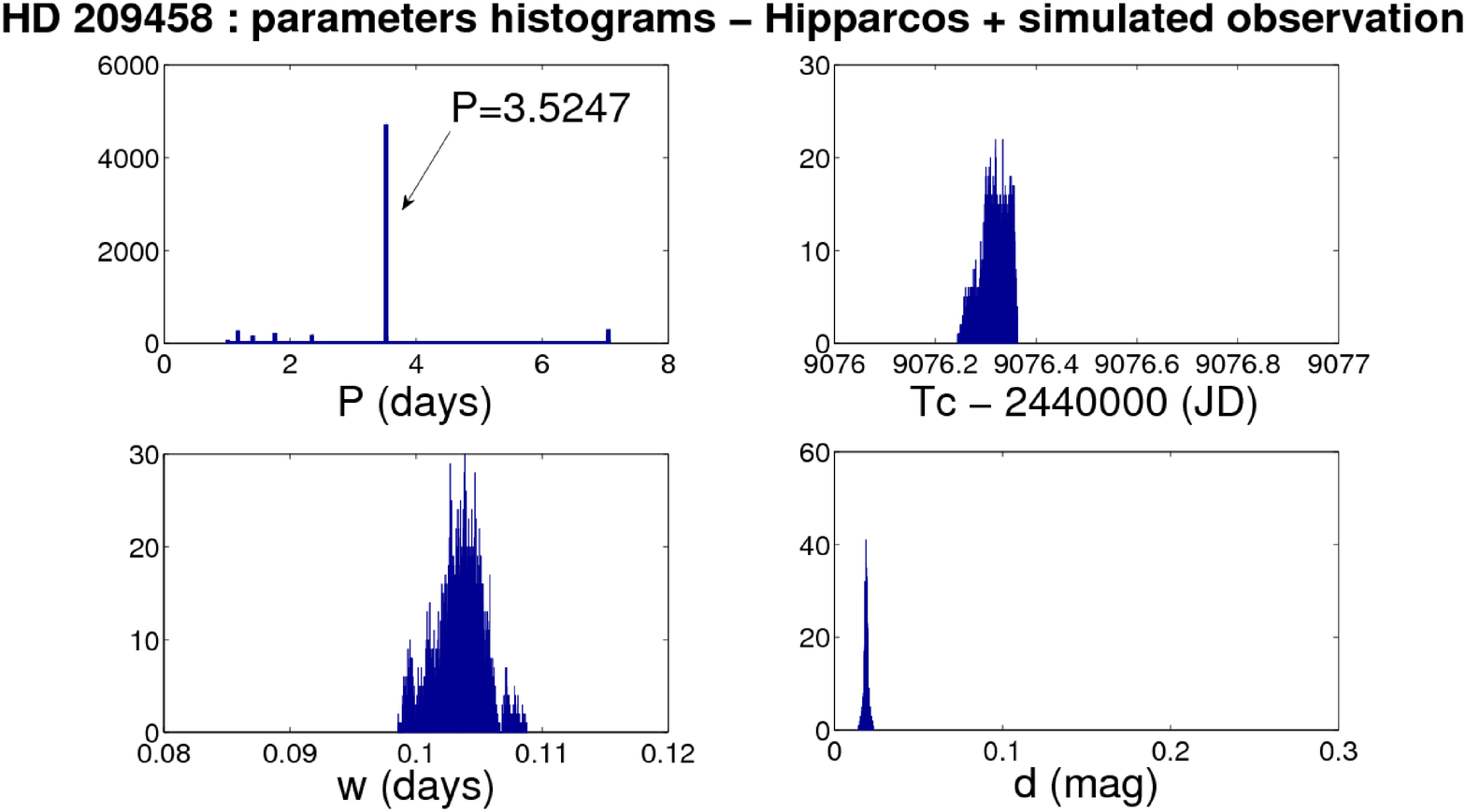}}
\subfigure{
\includegraphics[width=0.5\textwidth]{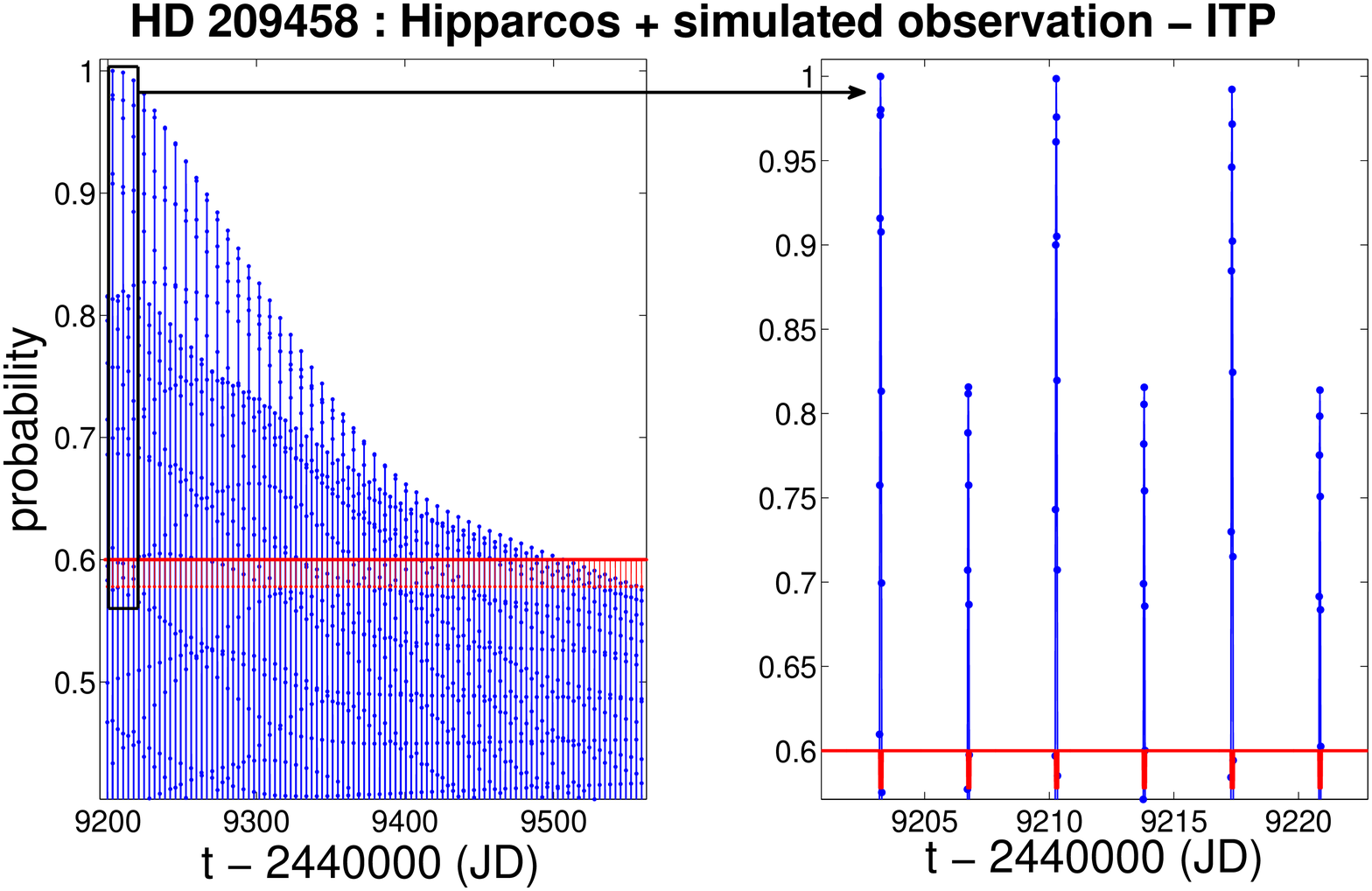}}
\caption{HD 209458: combined data sets of \textit{Hipparcos} and the simulated observation. The simulation generated four single measurements at the directed time by the ITP most significant peak. According to the known light curve of the transit this peak fits mid-transit time. Top: Histograms of the orbital parameters found using the MH algorithm for the combination of the data sets. Bottom: ITP for the first year after the simulation, compared with the known light curve of the transit. The transit detection was feasible, since all histograms are centered around the values of the orbital elements of the transiting planet: $P=3.5247$ d, $w=0.1$ d and $d=0.022$ mag \citep{2000ApJ...532L..51C}.}
\label{fig.HD209458_plus_sim_hist}
%\end{center}
\end{figure}

\begin{table*}
\centering
  \begin{minipage}{160mm}
\caption{Wald statistic of the transit depth posterior probability distribution function and Skewness of the ITP}
\label{table.waldvsSkewness}

    \begin{tabular}{lcc} \hline \hline              
      Data/simulation      & Wald statistic of the transit depth   & Skewness of the ITP   \\ \hline 
First simulation of transit & 5.5 & 5.5  \\
Second simulation of transit & 5.7 & 5.6 \\
HD 209458 - \textit{Hipparcos}  & 4.2 & 1.4 \\
HD 189733 - \textit{Hipparcos} & 3.6 & 4.7 \\
HD 209458 - \textit{Hipparcos} and one simulated observation & 18.1 & 4.8\\
HD 189733 - \textit{Hipparcos} and one simulated observation & 2.9 & 2.5\\
HD 189733 - \textit{Hipparcos} and two simulated observations & 3.4 & 2.7\\
HD 189733 - \textit{Hipparcos} and three simulated observations & 1.7 & 2.3\\
HD 189733 - \textit{Hipparcos} and four simulated observations & 4.3 & 2.2\\
HD 189733 - \textit{Hipparcos} and five simulated observations & 21.1 & 7.7\\
Noise alone & 1.3 & 0.07   \\
First permutation of transit simulation  & 1.5 & 0.87  \\
Second permutation of transit simulation & 2.2 & 0.001 \\
Third permutation of transit simulation & 2.1 & 0.6 \\
HD 209458 - \textit{Hipparcos} permutation & 1.4 & 0.5 \\
HD 189733 - \textit{Hipparcos} permutation & 1.5 & 1.0 \\
HD 86081 (no transit) - \textit{Hipparcos} & 1.3 & 0.47 \\
HD 212301 (no transit) - \textit{Hipparcos} & 1.4 & 0.43 \\
    \end{tabular}
 
\end{minipage}
\end{table*}

The simulation generated four observations (four single data points) inside and outside the
predicted time of the transit (the significant peak of the ITP), with a typical error of $0.001$
mag. 
In the simulation we used the known transit light curve to generate the observation.
The new histograms for the combined data sets are presented
on the top panel of Fig.~\ref{fig.HD209458_plus_sim_hist}. It is clear that
the first observation that could have been preformed using the
directed follow-up would have been enough to detect the transit, since
the histograms are centered around the parameters of the planetary
transit of HD 209458b. The simulated new observation exclude all the
spurious periods which the MH algorithm proposed based on the
\textit{Hipparcos} data alone. The Wald statistic for the transit depth now
increased to $W= 13.46$.  As can be seen from
Fig.~\ref{fig.HD209458_plus_sim_hist} (bottom panel), the new directed
follow-up that relies on both \textit{Hipparcos} and the new simulated
observation fits perfectly with the planetary transit of HD 209458b,
and with a high ITP value, which is close to 1, and with ITP skewness of $S=4.8$.
\section{HD 189733}\label{HD189733_5pars}

We next applied the procedure to the \textit{Hipparcos} data for HD
189733. \textit{Hipparcos} observed HD 189733 (HIP 98505) in non-uniformly
distributed $185$ epochs, over a time span of $1083$ d. We chose to
use only $176$ measurements that were accepted by at least one of the
two data reduction consortia. The estimated standard errors of each
individual $H_p$ magnitude are around $0.012$ mag, which is of the
order of the transit depth, similarly to HD 209458.

\begin{figure}
%\begin{center}
\subfigure{
\includegraphics[width=0.5\textwidth]{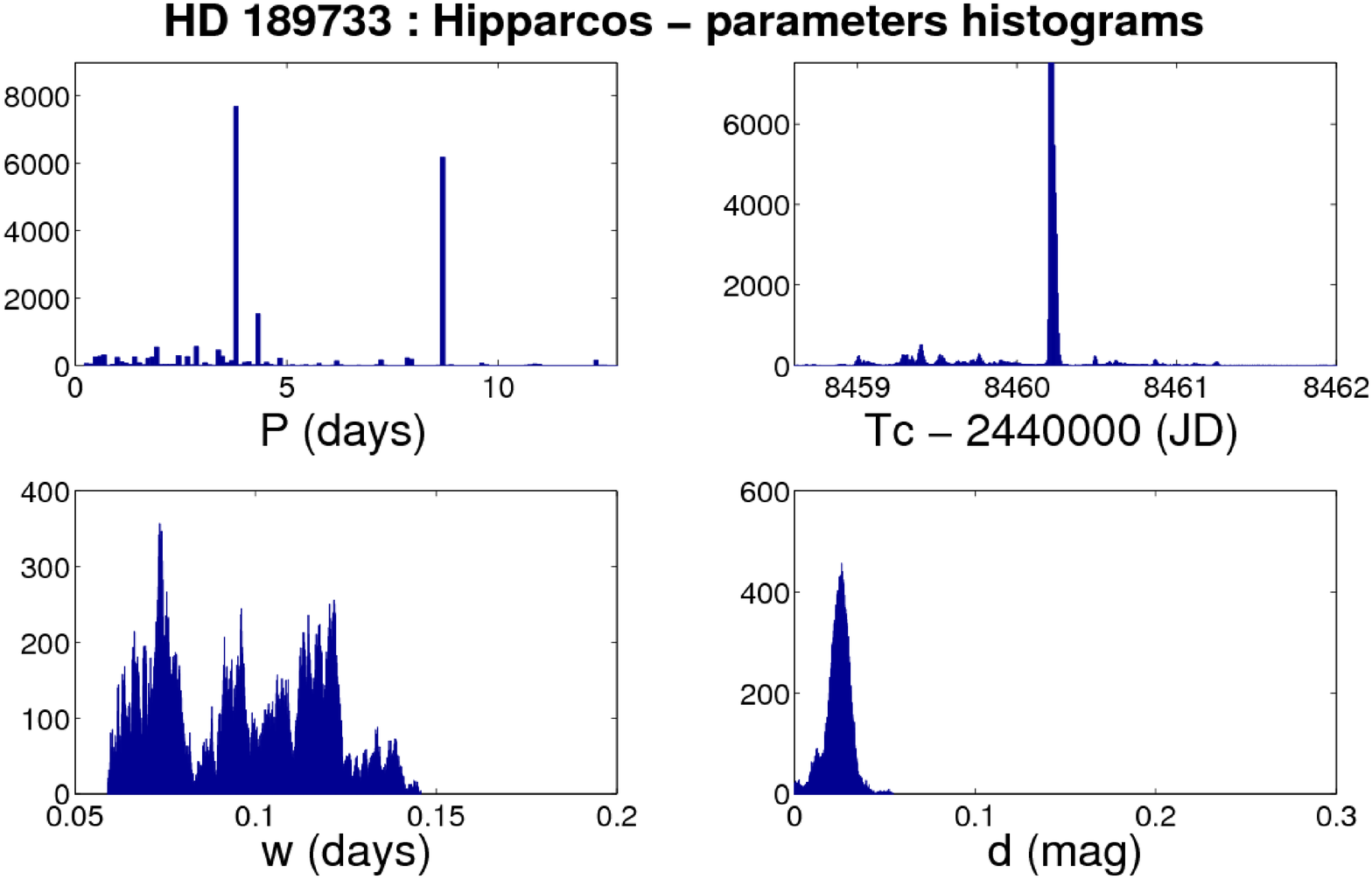}}
\subfigure{
\includegraphics[width=0.5\textwidth]{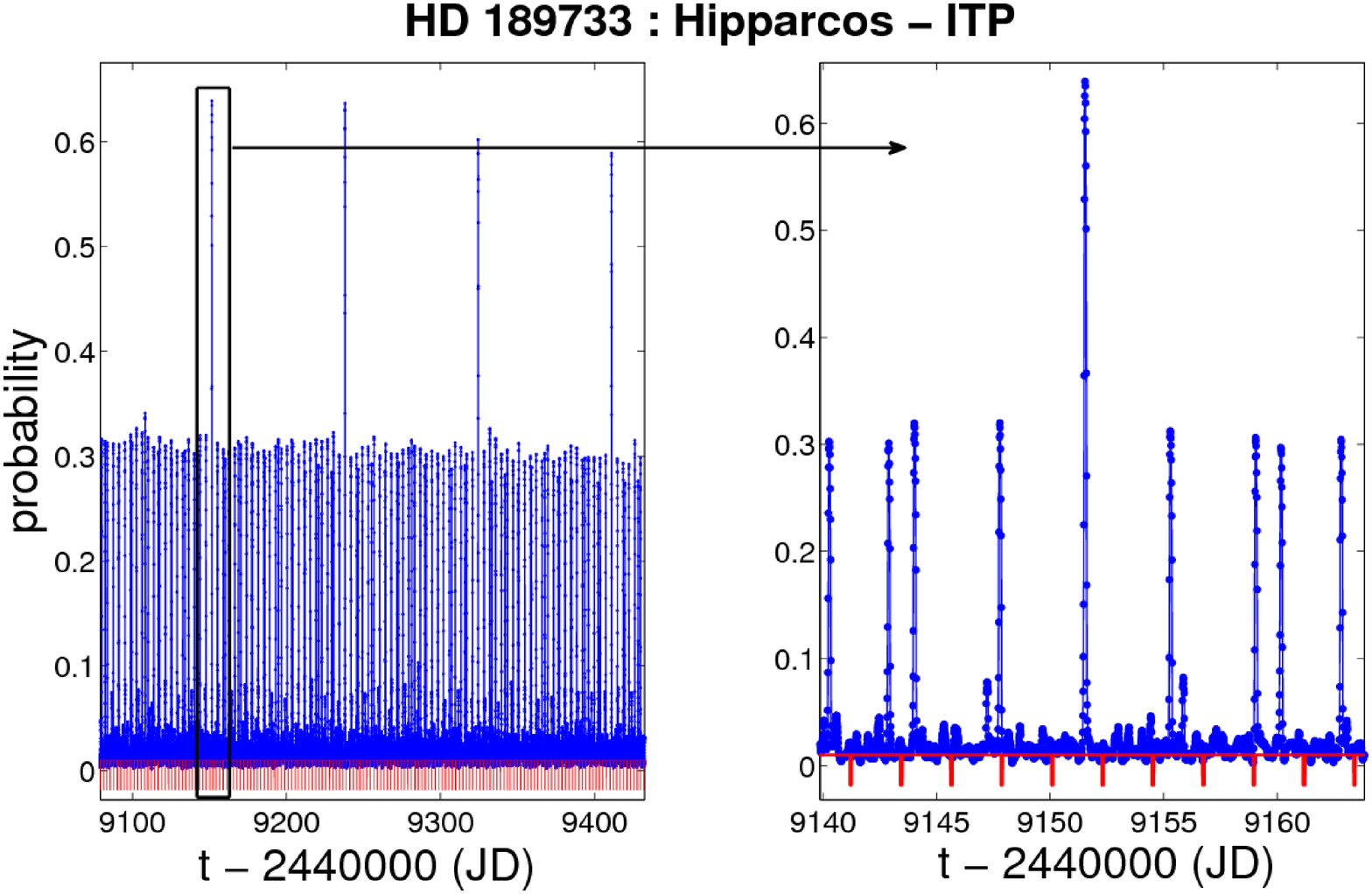}}
\caption{HD 189733. Top: Histograms of the orbital parameters found using the MH algorithm for the \textit{Hipparcos} data: the period, time of mid-transit, transit duration and the depth of the transit. The periods that are most probable using the MH algorithm differs from the planetary period ($P \sim 2.218$ d) \citep{2006A&A...445..341H}. Bottom: ITP for the first year after \textit{Hipparcos} observations, according to the procedure described in Section \ref{HD189733_5pars}, compared with the transit light curve, derived using the known orbital parameters \citep{2006A&A...445..341H}. The significant peaks do not fit mid-transit time, hence a single follow-up would not be sufficient for transit detection.}
\label{fig.HD189733hist}
%\end{center}
\end{figure}

The posterior distributions of the model parameters are shown on the
top panel of Fig.~\ref{fig.HD189733hist}. As can be seen from the
figure, the correct orbital period of the planet, $P=2.2185$ d,
\citep{2006A&A...445..341H} is not detected, and instead other periods fit the
data. The preferred mid-transit time is $T_c - 2440000 = 8460.21$ JD,
while the actual time of mid-transit according to Bouchy et al. (2005)
is $T_c-2440000 = 8460.11$ JD.

Again, we used the posterior distribution to perform the Wald test, to
test the hypothesis of the presence of a planet that is transiting the
star, although we obviously did not detect the correct period. The
expected value of the transit depth posterior distribution is $E(d)=
0.024$ mag, and the value of the Wald test is $W=
3.63$, which indicates that follow-up observations are worthwhile
since a transit is highly probable for this star.

\begin{figure}
%\begin{center}
\subfigure{
\includegraphics[width=0.5\textwidth]{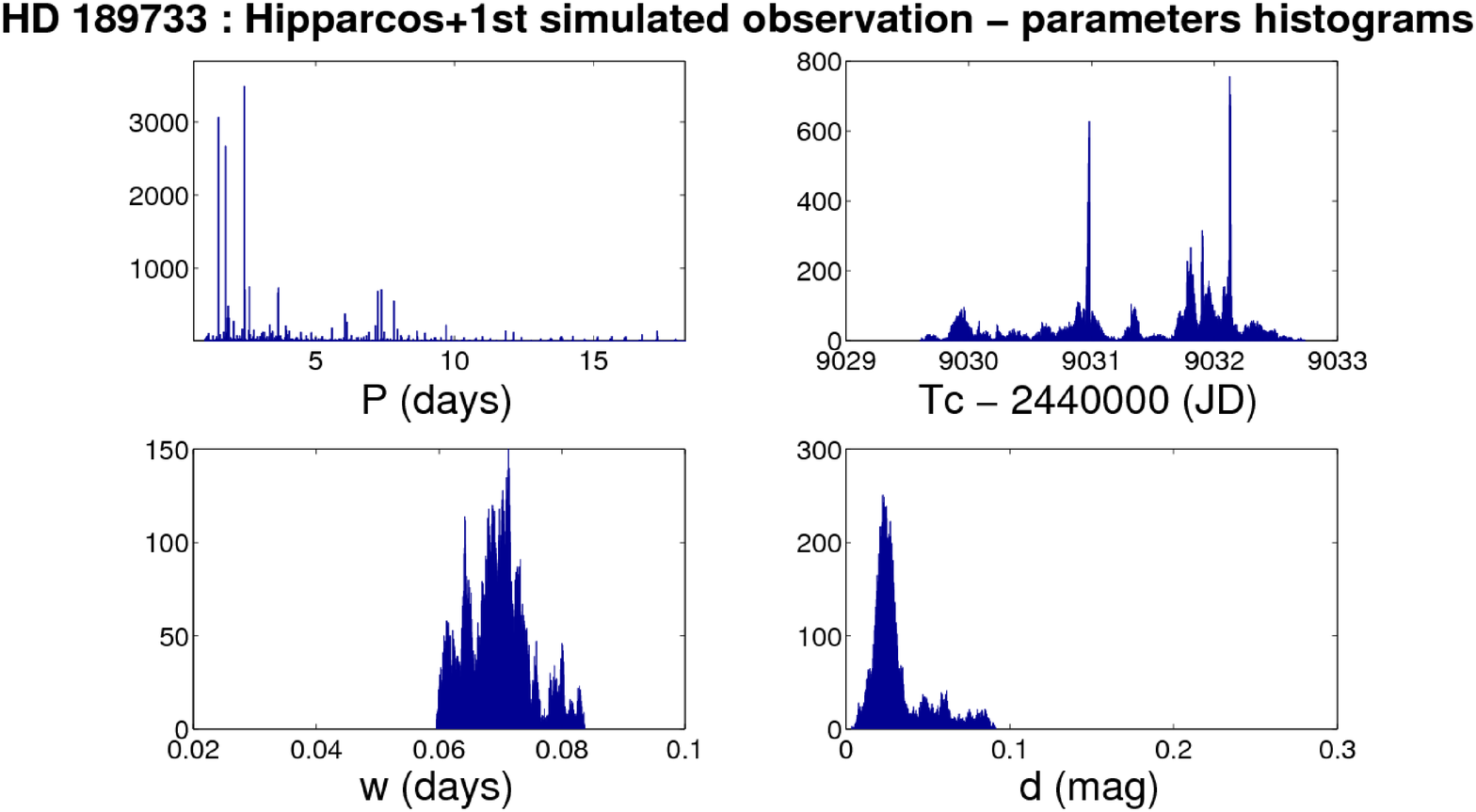}}
\subfigure{
\includegraphics[width=0.5\textwidth]{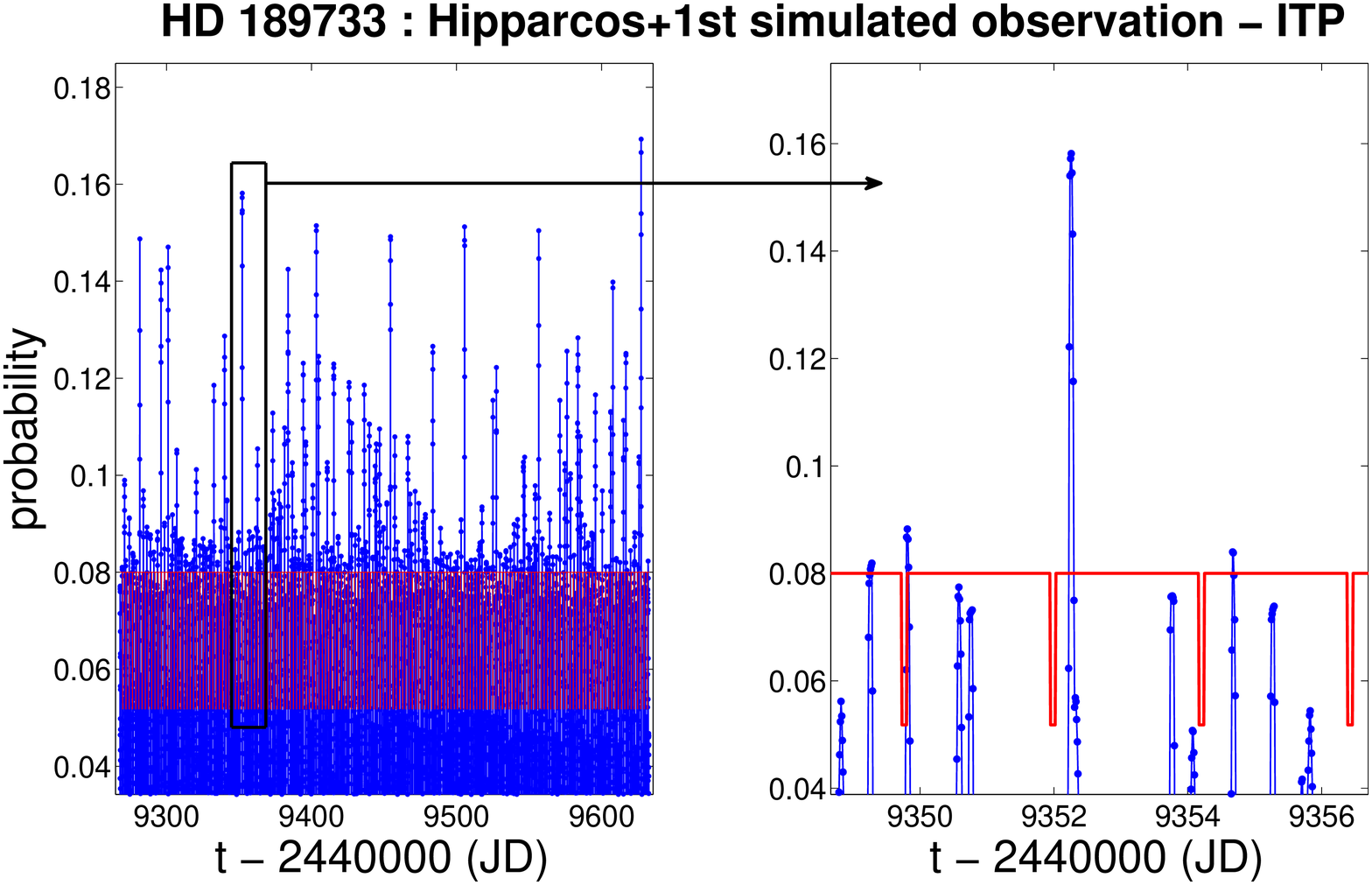}}
\caption{HD 189733: Combined data sets of \textit{Hipparcos} and the first observation simulation at the directed time. Top: Histograms of the orbital parameters. Bottom: ITP function.}
\label{fig.1}
%\end{center}
\end{figure}

The follow-up predictions we have for a year that starts one month
after the \textit{Hipparcos} observations are shown on the bottom panel of
Fig.~\ref{fig.HD189733hist}, compared with the known transit
light curve of HD 189733b. This time a single follow-up observation
would not have been enough to detect the transit, since the significant
peaks in the follow-up do not match the mid-transit time.  These
results for HD 189733 might have been caused by the star microvaribility
which \cite{2006A&A...445..341H} described. In their posterior
detection of HD 189733b in \textit{Hipparcos} they checked for long term
periodicity, and found several significant periods, that when removed
improved the $\chi^2$ statistic they got for the known orbital period
of the planet.  From the high ITP of the follow-up predictions, its skewness ($S=4.7$),
and the result of the Wald test, this star would get high priority for
follow-up observations, and although the observing time would not fit
at the middle of the transit, as soon as a new observation in the
preferred time would have been obtained, it would be added to the
previous data we already have, and the procedure would be repeated,
this time hopefully excluding the false periods.

\begin{figure}
%\begin{center}
\subfigure{
\includegraphics[width=0.5\textwidth]{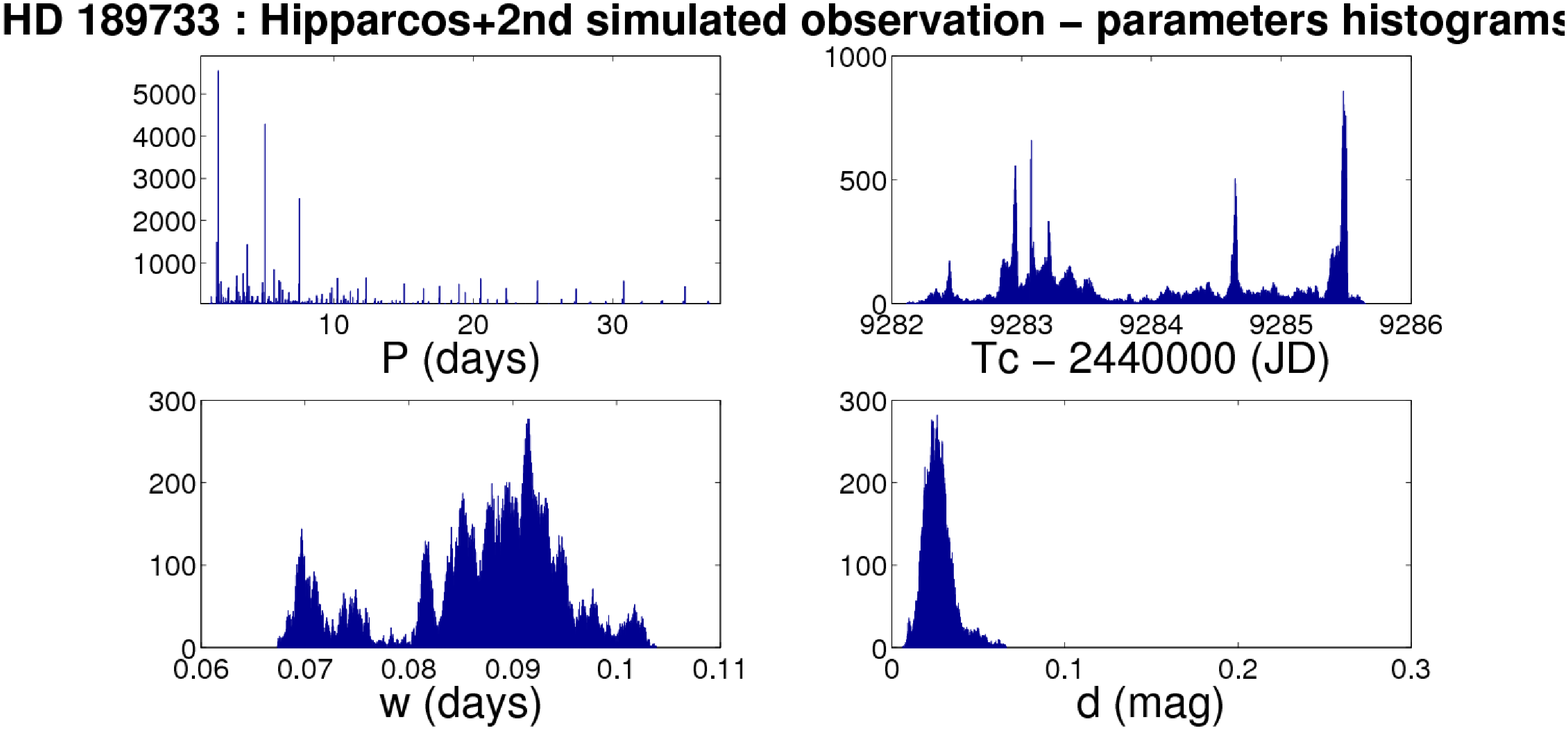}}
\subfigure{
\includegraphics[width=0.5\textwidth]{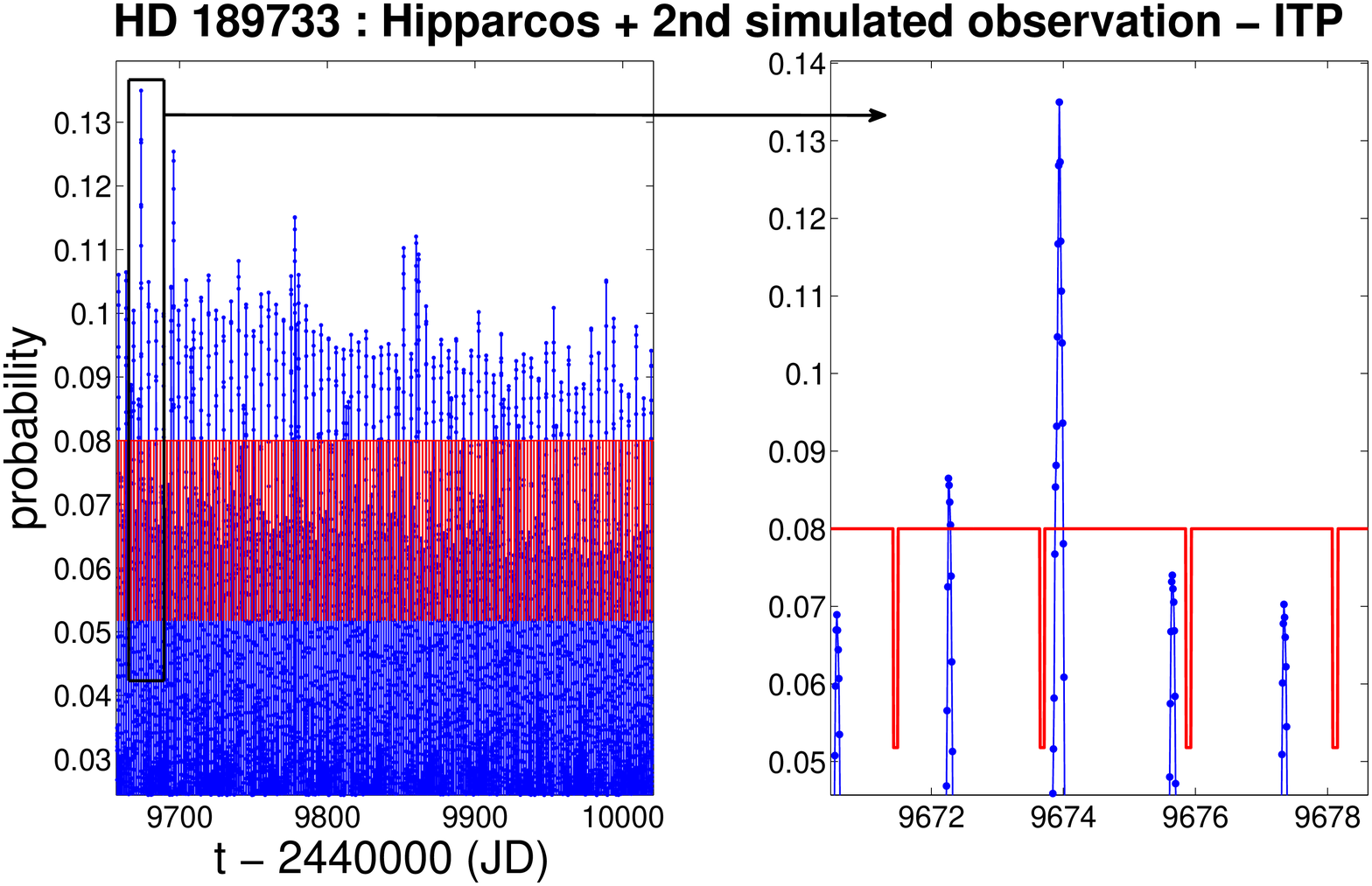}}
\caption{HD 189733 : Combined data sets of \textit{Hipparcos}, first and second observation simulations. Histograms of the orbital parameters and ITP function.}
\label{fig.2}
%\end{center}
\end{figure}

\begin{figure}
%\begin{center}
\subfigure{
\includegraphics[width=0.5\textwidth]{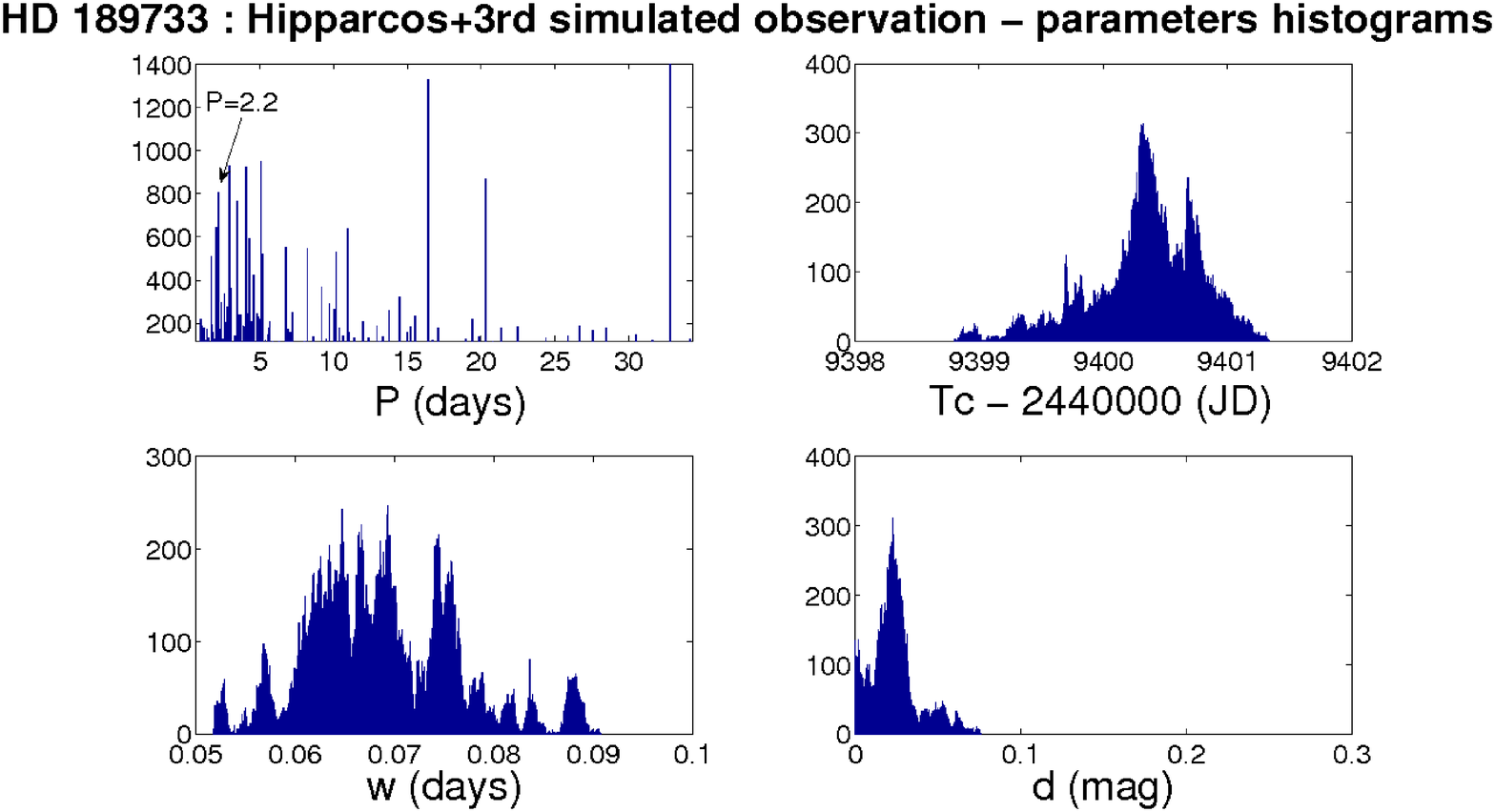}}
\subfigure{
\includegraphics[width=0.5\textwidth]{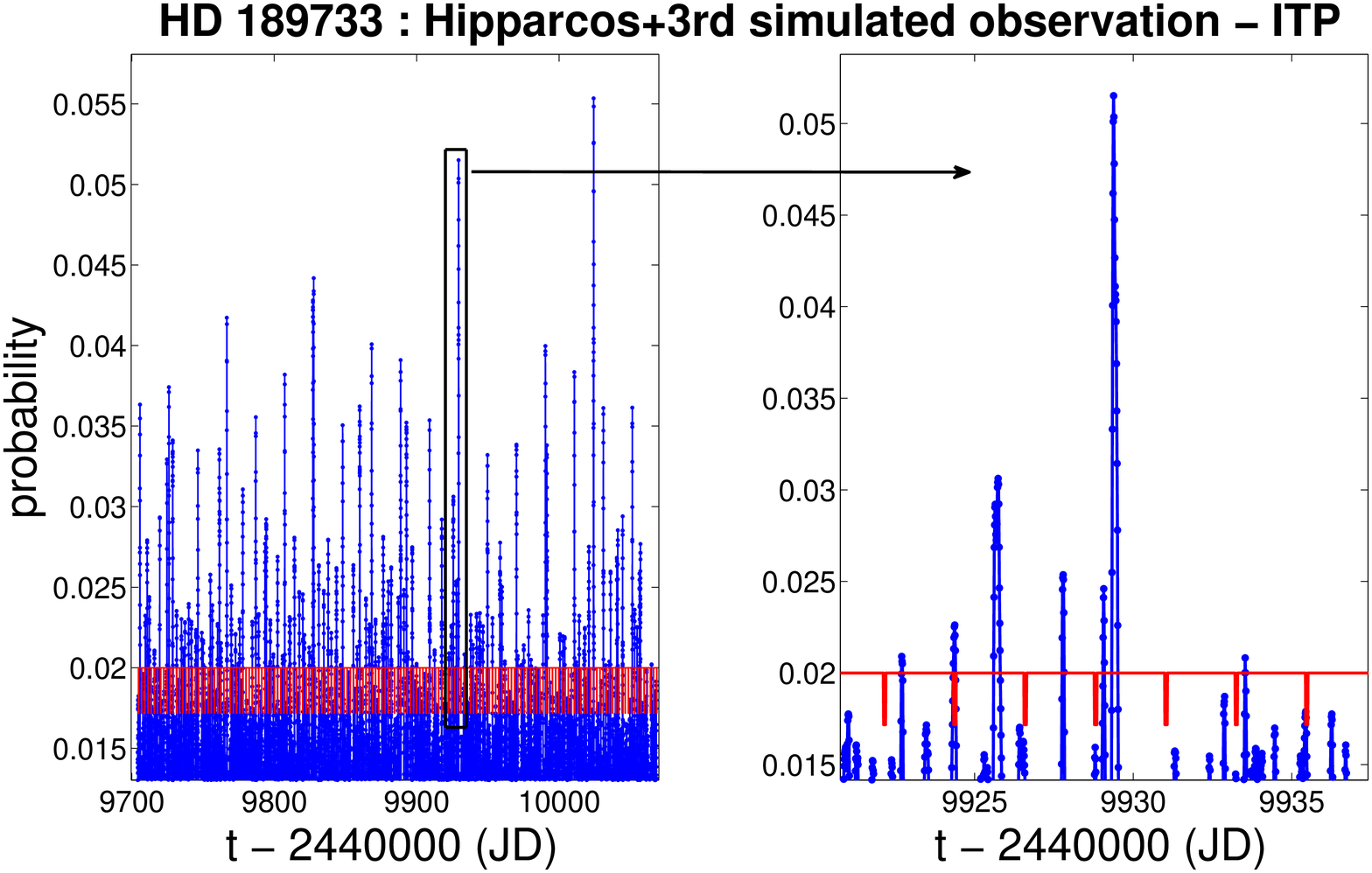}}
\caption{HD 189733 : Combined data sets of \textit{Hipparcos}, first, second and third observation simulations. Histograms of the orbital parameters and ITP function.}
\label{fig.3}
%\end{center}
\end{figure}

\begin{figure}
%\begin{center}
\subfigure{
\includegraphics[width=0.5\textwidth]{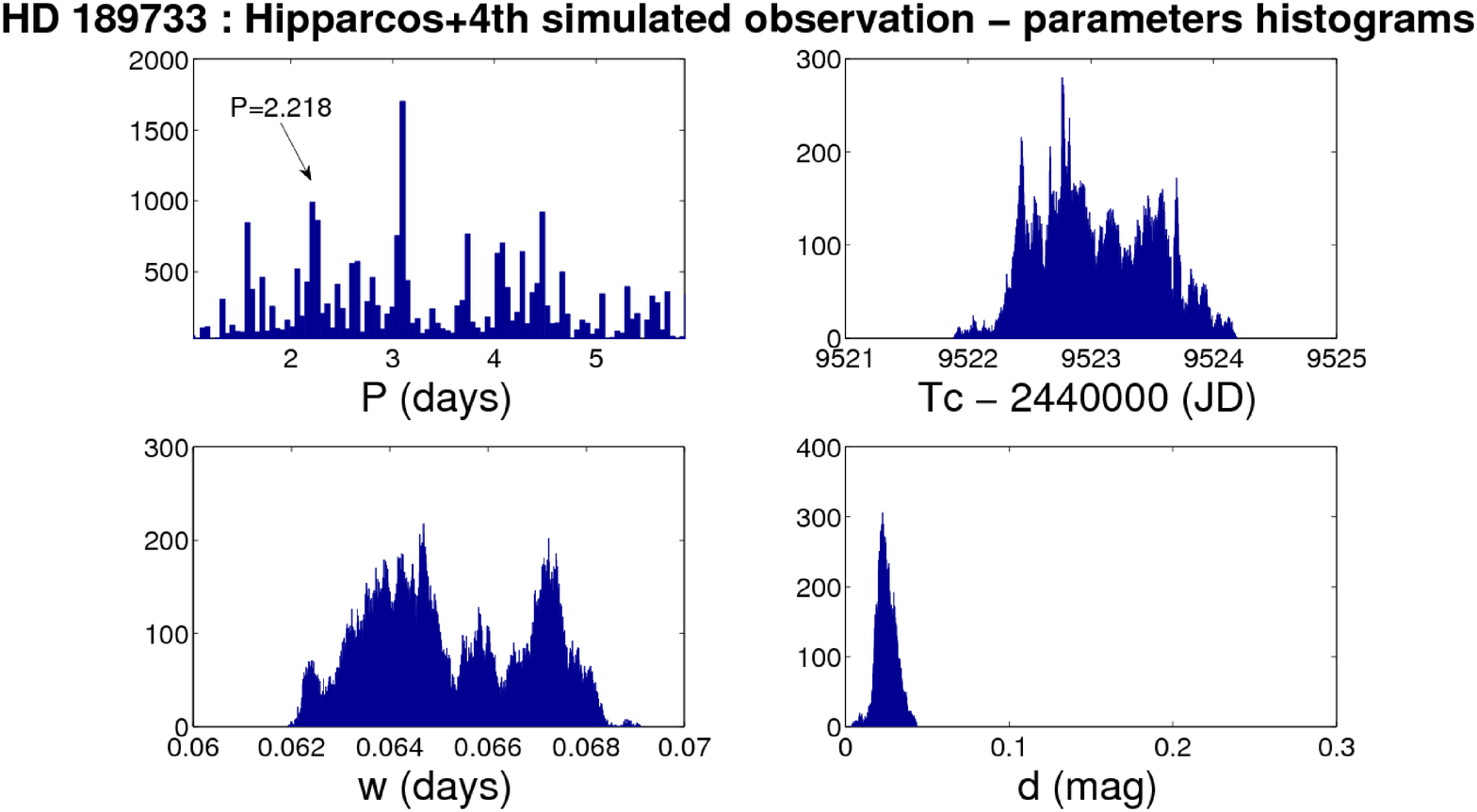}}
\subfigure{
\includegraphics[width=0.5\textwidth]{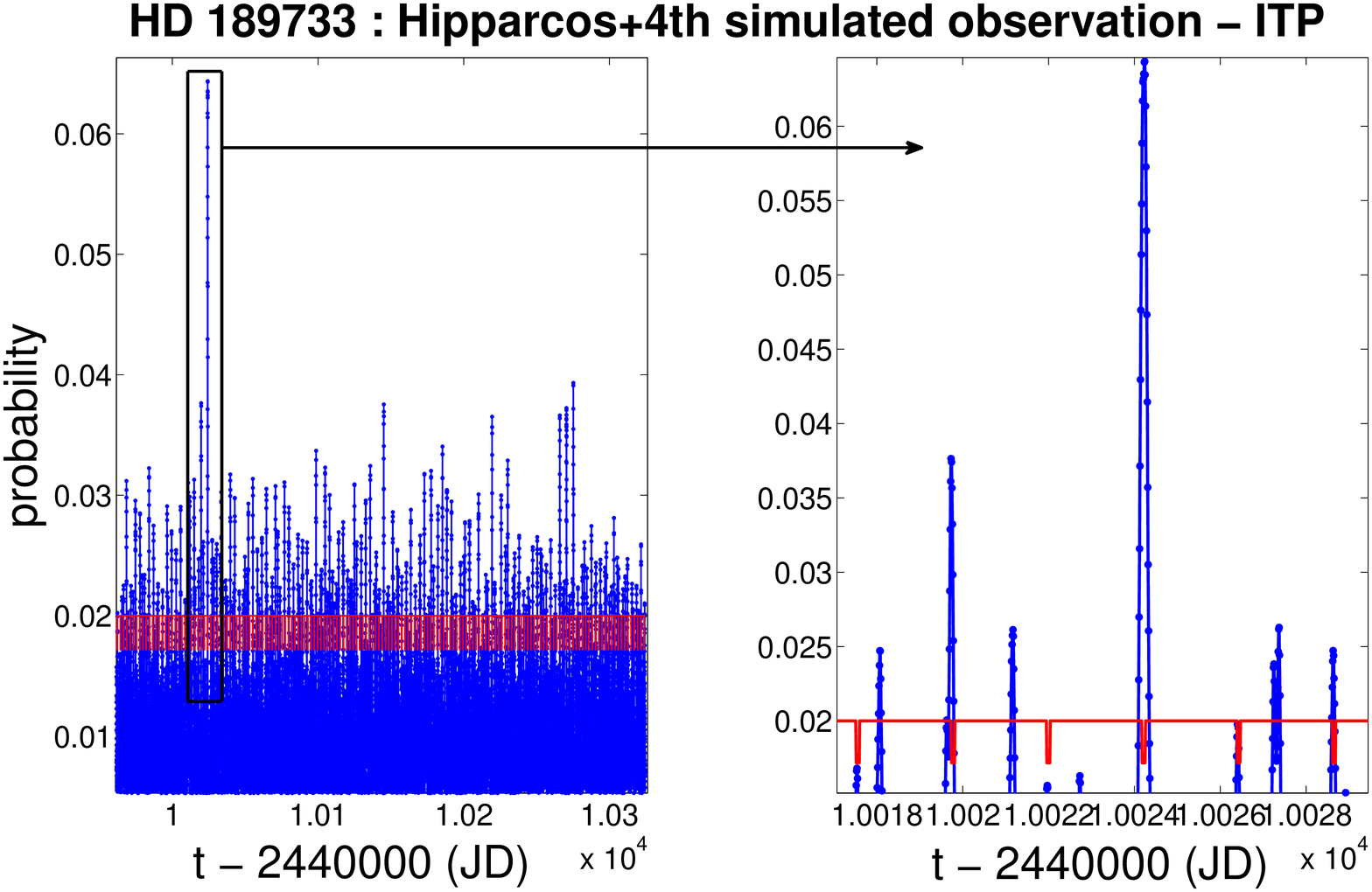}}
\caption{HD 189733 : Combined data sets of \textit{Hipparcos}, first, second, third and fourth observation simulations. Histograms of the orbital parameters and ITP function. This time the actual period of the planet ($P=2.218574$ d) is detected, and the most significant peak of the ITP fits the transit epoch.}
\label{fig.4}
%\end{center}
\end{figure}

\begin{figure}
%\begin{center}
\subfigure{
\includegraphics[width=0.5\textwidth]{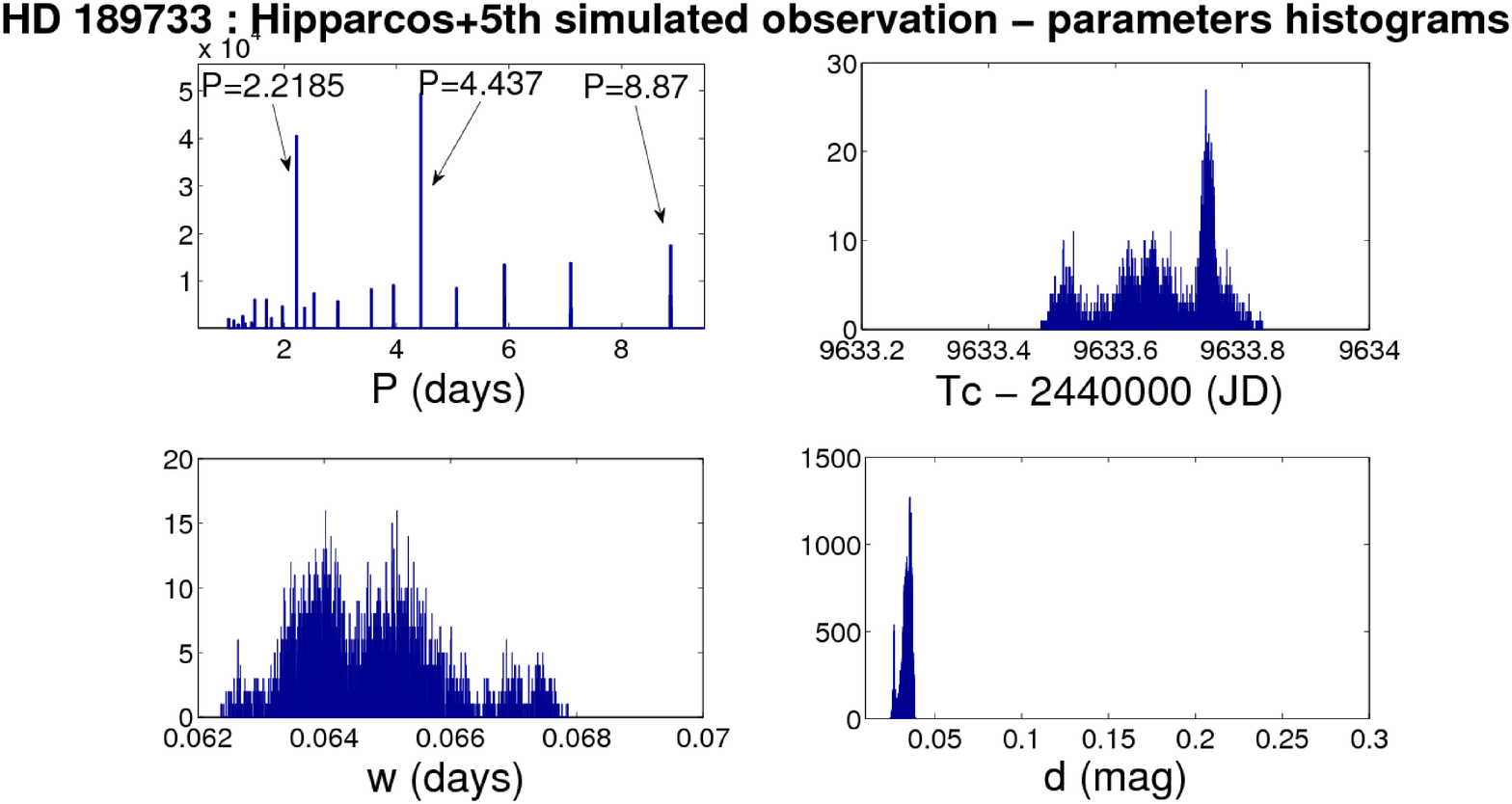}}
\subfigure{
\includegraphics[width=0.5\textwidth]{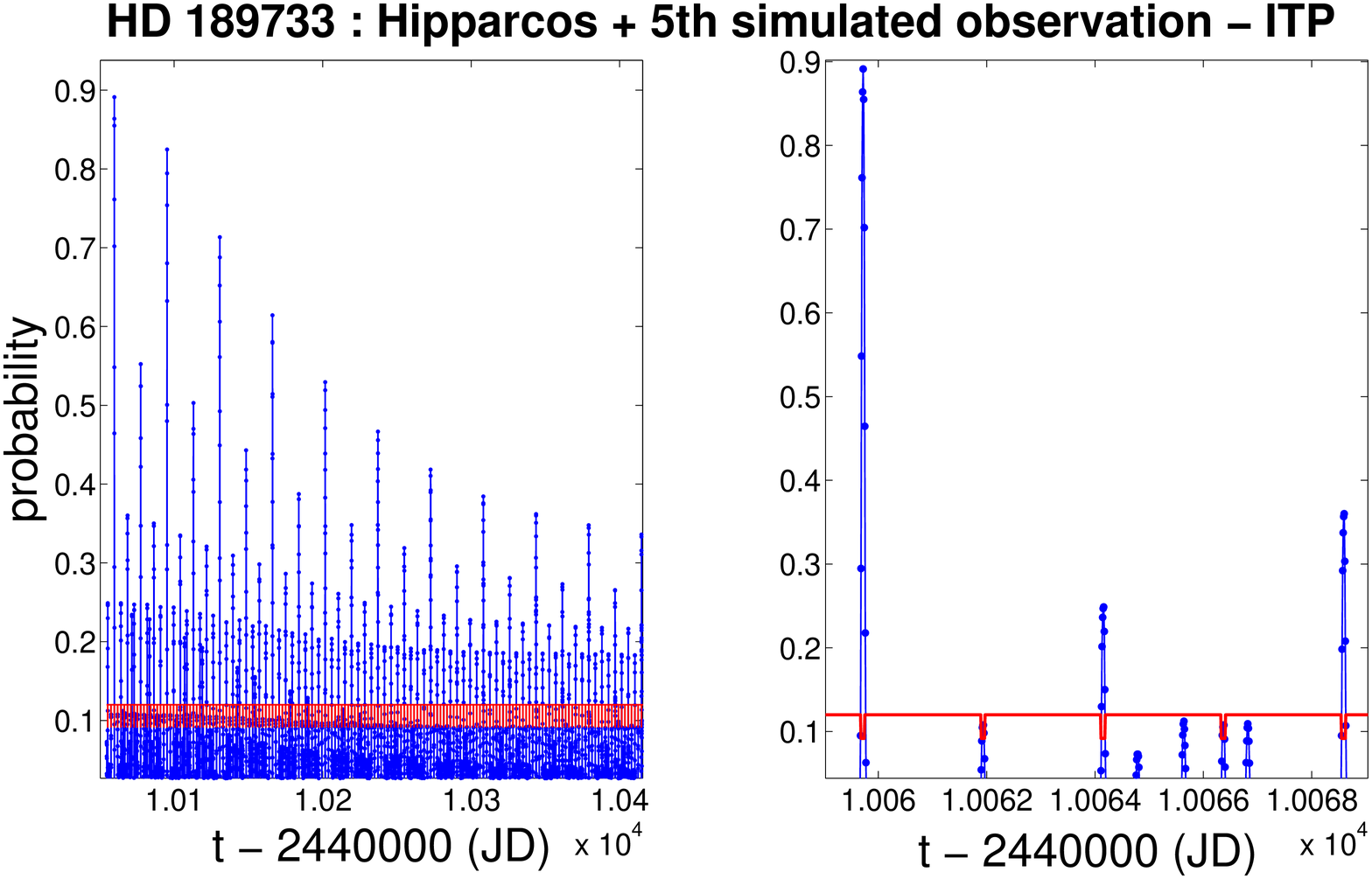}}
\caption{HD 189733: Combined data sets of \textit{Hipparcos}, and all five observation simulations. Histograms of the orbital parameters and ITP predictions. Using the MH algorithm, the transit is found, along with the parameters that characterize it ($P=2.218574$ d, $Tc=2453988.80331$ (HJD), $w=0.0589$ d, and $d=0.033$ mag in the \textit{Hipparcos} $Hp$ system, derived by \citet{2006A&A...445..341H} and \citet{2007AJ....133.1828W}). All the new peaks of the ITP fits mid-transit time with high detection probability.}
\label{fig.5}
%\end{center}
\end{figure}

In order to check this claim we simulated a follow-up observation at
the time the algorithm directed. Recall that in this case (as opposed
to the case of HD 209458), the directed time was not in transit. We
then added it to the \textit{Hipparcos} observation to recalculate the
parameters posterior distributions, as well as to propose a new time
for the next observation.  We had to simulate a total of five follow-up
observations, each containing four ``exposures'' with an error of
$0.001$ mag, in order to finally detect the transit itself. In
Figs~\ref{fig.1}-\ref{fig.5} we present the changes in the
posterior distribution, as well as the directed follow-up predictions,
as more simulated observations are added to the original data. Each
observation eliminates some of the periods, making room for other
periods to emerge, until an actual transit is observed at the final
observation. By the fifth follow-up observation, the only probable
periods left were the transit period and its multiples. Table \ref{table.waldvsSkewness} summarizes the Wald statistics and the skewness of the ITP for the original \textit{Hipparcos} data, together with the simulated observations.

\section{'Sanity Checks'}\label{sanity_check}

As we demonstrated above, it was possible to detect the transiting
exoplanets orbiting both HD 209458 and HD 189733 using the follow-up
strategy we proposed. We now want to demonstrate cases where no
transit signal is supposed to exist.

\subsection{HD 209458 - permuted data}\label{HD209458_perm}

We perform the first test by randomly permuting the \textit{Hipparcos} data of
HD 209458 and repeating the simulated follow-up procedure described
above.  The result of the Wald test, $W= 1.45$, shows that the
likelihood of a transit for the permuted data is low.  The predicted
ITP for the directed follow-up is small for all the
time-span that we checked, with a skewness value of $S=0.5$, which again implied that there were no
preferred time to observe a transit. This means that it is unlikely
that there is a transit signal in the permuted data.

\subsection{HD 189733 - permuted data}\label{HD189733_perm}

\begin{figure}
%\begin{center}
\subfigure{
\includegraphics[width=0.5\textwidth]{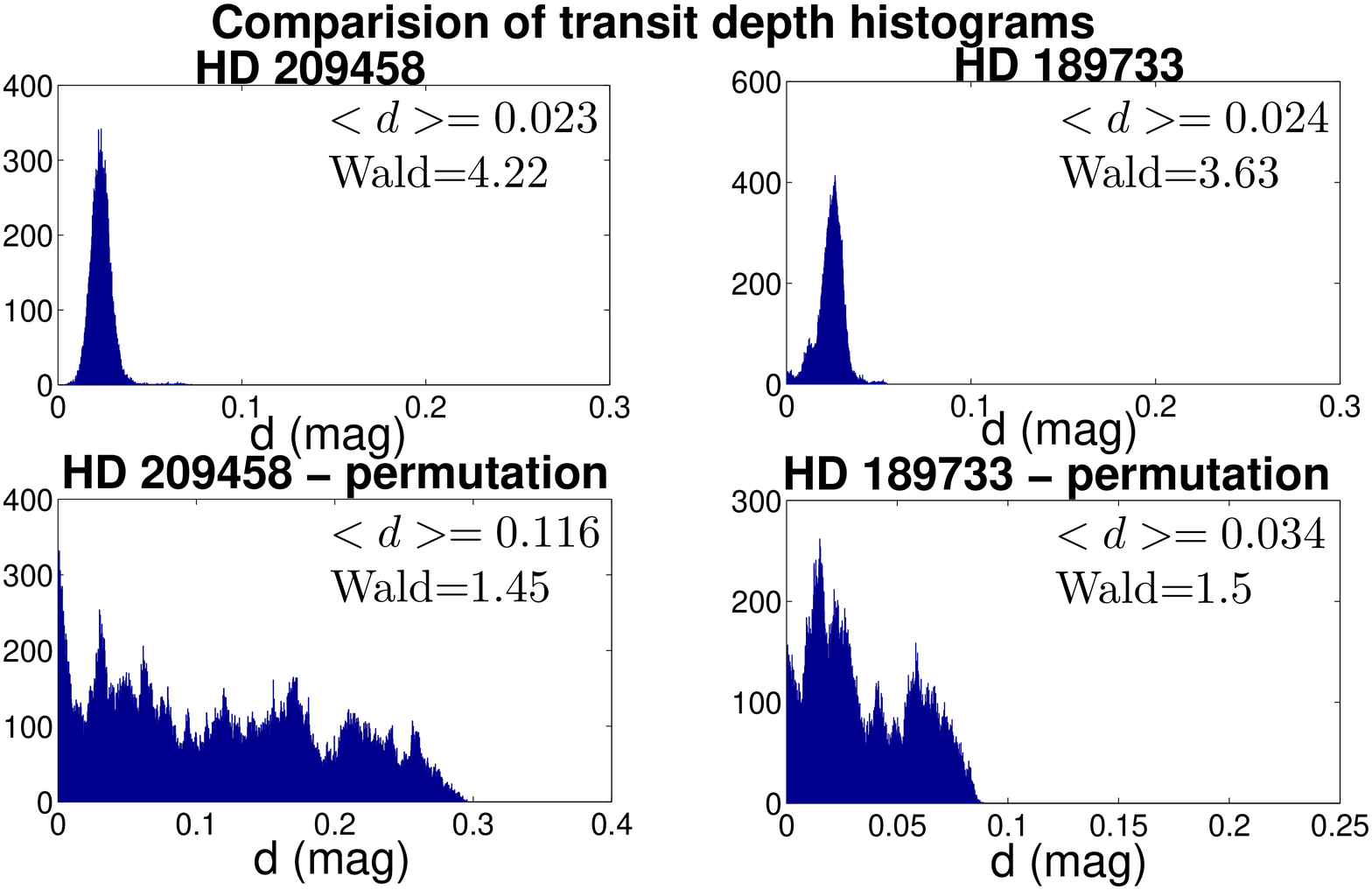}}
\subfigure{
\includegraphics[width=0.5\textwidth]{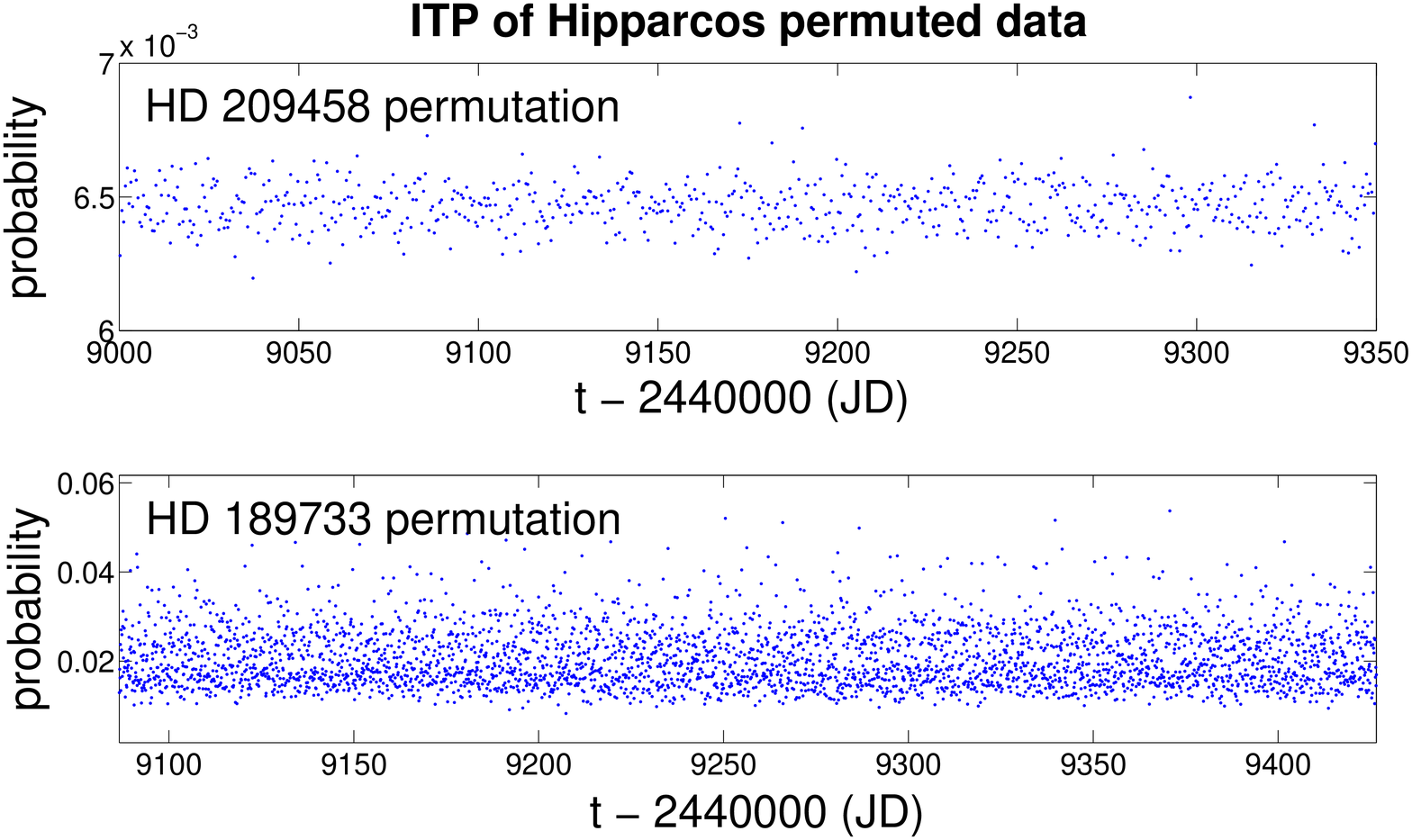}}
\caption{HD 209458 and HD 189733. Top: Comparison between the transit-depth histograms for the \textit{Hipparcos} data and for the permuted data, with the associated values of the Wald test. Bottom: ITP function for the permuted data sets.}
\label{fig.depthhistHD209458HD189733}
%\end{center}
\end{figure}

As for HD 209458, we randomly permuted the data of HD 189733 in order
to test our procedure. The Wald statistic for the permuted data is
$W=1.5$, which means that it is not likely to find a transit based on
the permuted data.  Although not as low as for the permuted data of HD
209458 the computed ITP in the directed follow-up, its skewness ($S=1.0$),
combined with the small value of the Wald statistic, means that it is
not worthwhile to preform follow-up observations in search of
transits. Fig.~\ref{fig.depthhistHD209458HD189733} shows the comparison between the 
transit-depth histograms of HD 189733 and HD 209458 and the histograms of the permuted 
data of both stars, combined
with the follow-up predictions of the permuted data. The clear
difference between the transit depth histograms of data that contain
a transit signal and the permuted data, as expressed by the
Wald statistic, is a good indication for the Wald statistic strength
as a prioritization tool for follow-up observations.

\subsection{HD 86081 and HD 212301}\label{HD86081}

\begin{figure}
%\begin{center}
\subfigure{
\includegraphics[width=0.5\textwidth]{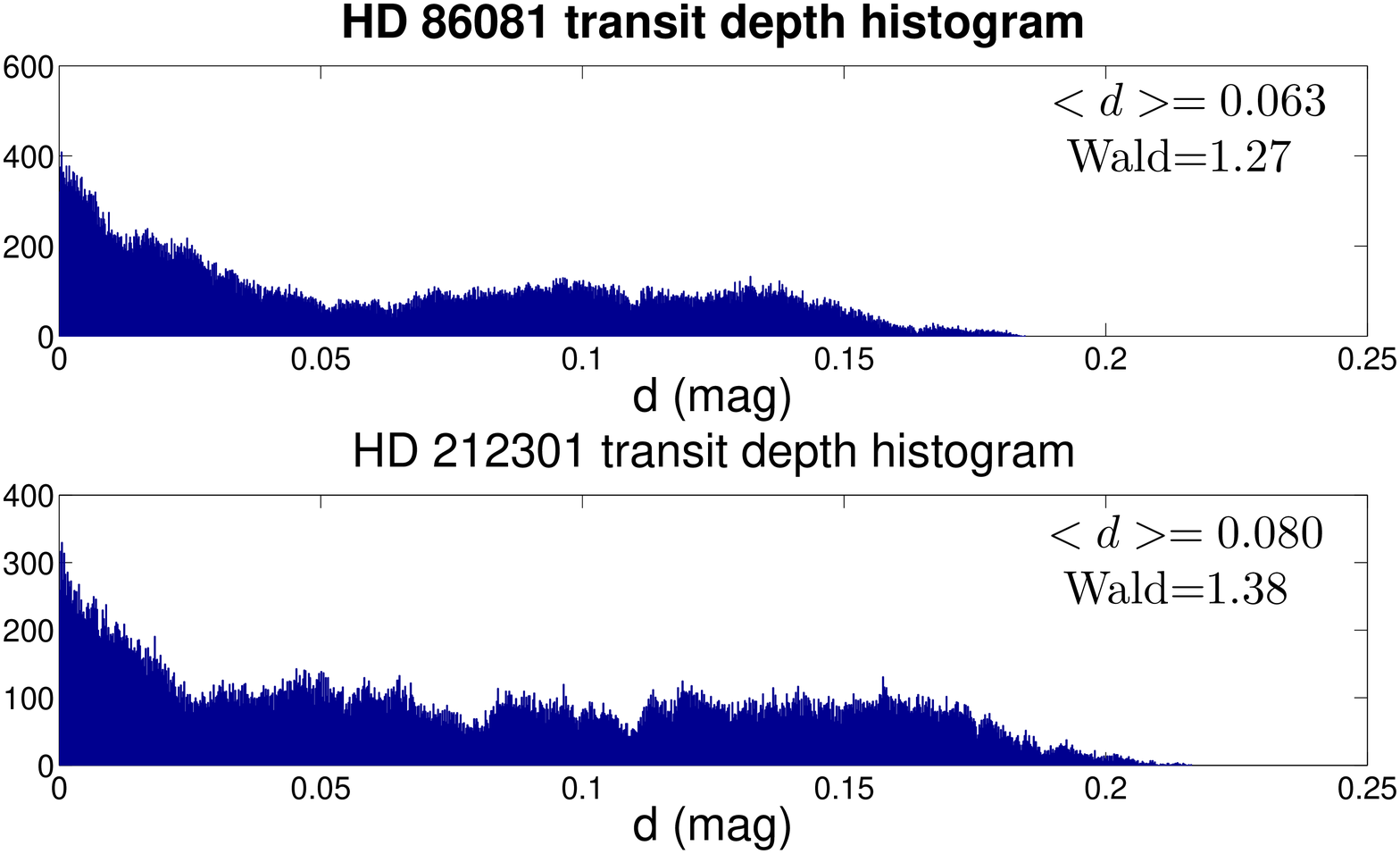}}
\subfigure{
\includegraphics[width=0.5\textwidth]{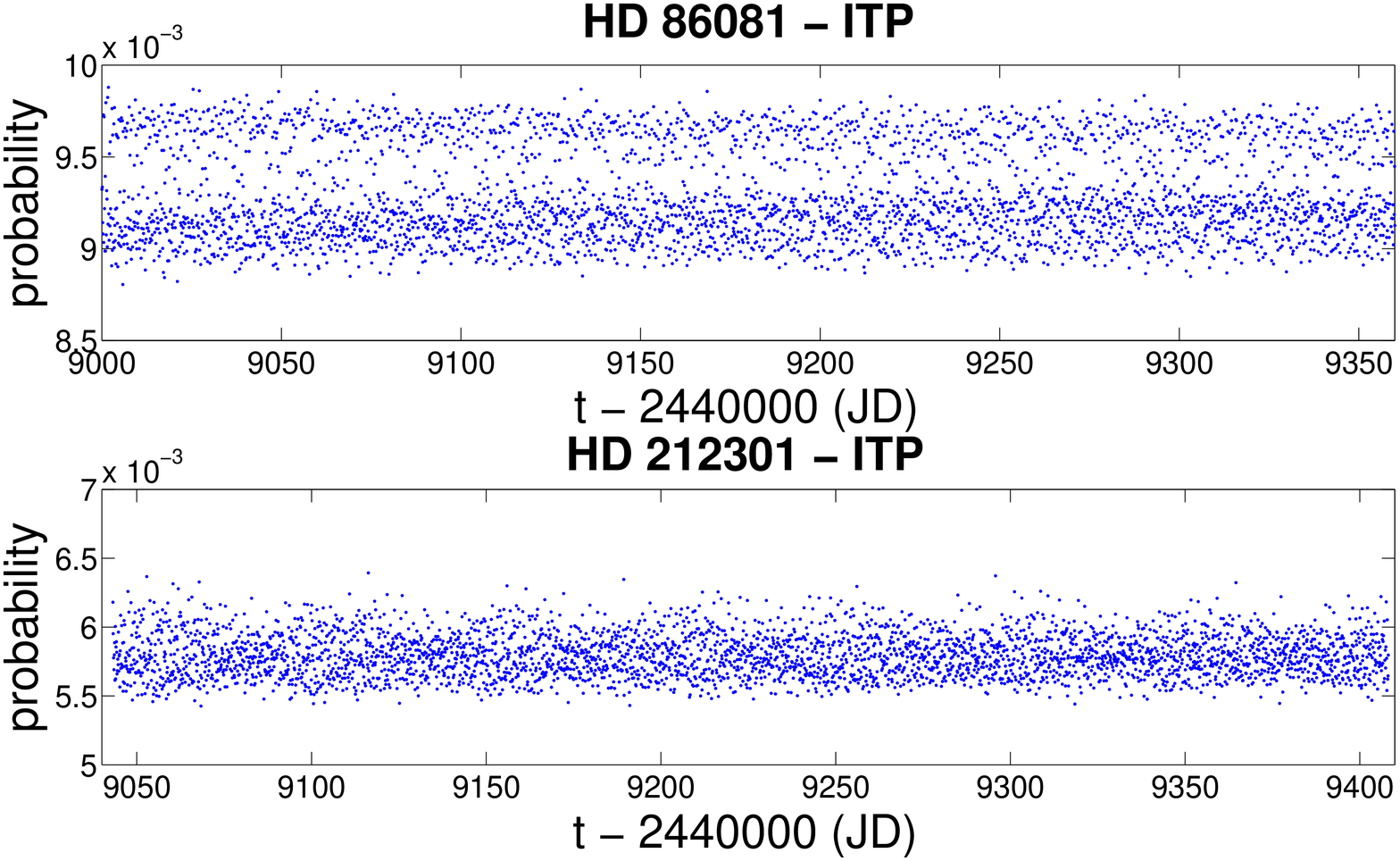}}
\caption{HD 86081 and HD 212301: transit-depth histograms and ITP predictions as derived using \textit{Hipparcos} data for both stars. The depth histograms are distributed over the whole range, with low Wald statistic for the transit depth. Together with the low significance of the ITP, the star would not be prioritized for follow-up observations.}
\label{fig.depthhistHD86081HD212301}
%\end{center}
\end{figure}

Besides examining the randomly permuted data of HD 209458 and HD
189733, we also applied our procedure on two stars for which we have
reasons to believe there is no transit signal.  We chose the two stars
HD 86081 and HD 212301, which are known to harbor short-period
planets.  Since the planets are known to have short orbital periods,
the fact that no transits were detected \citep[][]{2006ApJ...647..600J, 2006A&A...451..345L}, means it is very unlikely
that the stars have other Hot Jupiters orbiting them, thus making them
perfect targets for testing our procedure, as negative test cases.
\textit{Hipparcos} observed HD 86081 for $71$ reliable epochs, and HD 212301
for $123$, during the operation time of the satellite, numbers which are similar to the number of Hipparcos measurements of HD 209458.  We applied the
strategy on both data sets. Fig.~\ref{fig.depthhistHD86081HD212301}
shows the corresponding depth histograms and the ITP function for both stars.
HD 86081 and HD 212301 did not show any significant value for the
ITP, and the highest value was at least one order of
magnitude below the predictions for HD 209458 and HD 189733. As a result, the skewness values of the ITP of both stars were low as well, (smaller then $0.5$) which indicates that a follow-up observation is not worthwhile for those stars. The transit-depth Wald statistics for the two stars were $W= 1.2$ for HD
86081 and $W=1.4$ for HD 212301, which again indicates that transits
are unlikely.

\section{concluding remarks}\label{cuncluding}

In this paper we proposed a novel approach to the design of follow-up
observations of low-cadence photometric surveys, in a way that will
maximize the chances to detect planetary transits. Examples of such
surveys are \textit{Hipparcos}, \textit{ASAS}, and \textit{Gaia} as \textit{Hipparcos} successor.
The strategy may also be beneficial for the Large Synoptic Survey Telescope \citep{2011EAS....45..281J}, and for Pan-STARRS ground-based survey, especially for directing follow-up observations of hot Jupiters transiting M-dwarf stars in the Medium Deep survey \citep[][]{2009ApJ...704.1519D, 2008AIPC.1082..275F}.

We tested our proposed procedure on two stars with transiting planets
that were observed by \textit{Hipparcos} during transits: HD 209458 and HD
189733. We showed that without any prior information regarding the
orbital elements of the planets, it was possible to use the available
data base of \textit{Hipparcos} to direct follow-up observations for
both stars and thus detect the planetary transits in minimal
observational effort. This makes use of the fact that the Bayesian
approach allows the inclusion of new data, that reflect new state of
knowledge, in an easy and straightforward fashion.

The \textit{Hipparcos} examples we analyzed are only test cases to demonstrate
the algorithm capabilities. Using \textit{Hipparcos} in such fashion to detect
planets is already impractical, due to the long time that elapsed
since the completion of the mission.  The effect of the elapsing time
is clearly seen in the way the ITP decreases during
$10$ yr (Fig.~\ref{fig.HD209458followup10years}). We have shown
that one year after \textit{Hipparcos}, it was possible to use its data to
direct photometric follow-up observations that could have detected the
planetary transits in only one follow-up observation for HD 209458,
and in five observations for HD 189733.

In cases where only the Wald statistic has high significance, but the
ITP is relatively low, we might recommend
performing spectroscopic follow-up instead of photometric one, since
RV search is less dependent on precise knowledge of the
transit phase, because the goal is then to sample all phases of the
orbit. Since \textit{Hipparcos} observations were performed almost two decades
ago, and due to the fact that the ITP lost its
significance, it may be
more productive to perform RV follow-up observation for potential
stars found in \textit{Hipparcos} alone. We will explore this option in future work.

Obviously, the procedure suggested here should not only be confined to the
search for transiting planets, but can also be applied for searching
other kinds of periodic variables, such as eclipsing binaries and
Chepeids. This will probably require some modifications of the
procedure and algorithm. In this context, it is important to mention a
similar approach of adaptive scheduling, which Tom Loredo proposed for the purpose of optimizing RV observations.
The approach, adaptive
Bayesian exploration (ABE; \citet{2004AIPC..707..330L}), is much more general, attempting to optimize the
information obtained by every additional observation for the purpose
of estimating the parameters of the model behind the observations. The
formulation of our problem is much more specific and simple - we want
to optimize our chances to 'catch' the transit using well-scheduled follow-up observations. 
While every RV measurement contributes in some way to the orbital solution, the contribution of an individual photometric measurement to the transit solution boils down to the binary question whether it is in the transit or not. Thus, ABE uses an elaborate merit function that quantifies the amount of information in the RV measurements. ABE can probably be applied to our problem as well, but we feel it would be redundant due to the simpler nature of the problem. We speculate that the two approaches would yield very similar results.

Our experience shows that the MH algorithm and the ITP tend to find all possible periods that fit the
data. Because of the low cadence of \textit{Hipparcos} measurements, some hypothetical
models may fit the data simply because the ``transits'' occur during
'gap' intervals, when no observations were made.  Thus, the follow-up
prediction function, when generalized to a broader model space of
periodic variables, will allow constructing a follow-up strategy
that will complement the low-cadence observations in a way that will
optimize the period coverage.

MCMC methods, such as the MH algorithm, can be
very demanding in terms of processing time.  Therefore, improving the
efficiency and automatizing the strategy in order to explore large
data bases is crucial to its usefulness. Thus, we are examining the
idea of reducing the amount of model parameters that the MH algorithm
explores to three main parameters: the transit period, duration, and
mid-transit epoch, while marginalizing over the other two parameters:
the transit depth and mean magnitude out of transit. The
marginalization will hopefully shorten the computing time.  Another
idea worth examining is using a BLS-like algorithm, which will scan the ($P,T_c,w$) space and calculate the likelihood of each configuration, from which it will build, in a Bayesian fashion, the ITP function. Such scanning is obviously a compromise, since it is discrete and finite by nature, and the coverage of the parameter space may be lacking. However, the gain in computation time compared to a Monte Carlo approach might be worth the price.

At this stage the simulations we have presented in this paper are a
feasibility test, based on \textit{Hipparcos} Epoch Photometry.  The
encouraging preliminary results we present here lead us to believe
that the strategy can be beneficial for \textit{Hipparcos} successor, \textit{Gaia}
\citep{2009sf2a.conf...45E}. \textit{Gaia}, whose expected launch is planned to
2012, will measure about a billion stars in our Galaxy and in the
Local Group, and will perform , besides ultraprecise astrometry, also
spectral and photometric observations. \textit{Gaia} is supposed to improve on
the accuracy of \textit{Hipparcos} using larger mirrors, more efficient cameras
and detectors and better software to reduce the data.  In its
photometric mission, \textit{Gaia} will scan the whole sky, with a photometric
precision of 1 mmag for the brightest stars, and up to $20$ mmag at a
magnitude of $20$ \citep{2009sf2a.conf...45E}. \textit{Gaia} main exoplanets
search programme is focused on detection through astrometric motion
measurements. The strategy we propose here may be generalized to
direct follow-up efforts of \textit{Gaia}'s photometry, aimed to detect
transiting exoplanets.

\section*{Acknowledgments}

This research was supported by The Israel Science Foundation
and The Adler Foundation for Space Research (grant No. 119/07).

% 
% \bibliographystyle{mn2e}
% \bibliography{ms_proof.bib}

% 
% \begin{figure}
% \begin{center}
% \subfigure{
% \includegraphics[width=0.5\textwidth]{HD209458_5pars_random_hist_a_rev1.eps}}
% \subfigure{
% \includegraphics[width=0.5\textwidth]{HD209458_5pars_followup_zoom1_a_rev1.eps}}
% \caption{Top: HD 209458 - Histograms of the posterior probability distribution functions of four orbital parameters found using the MH algorithm for the \textit{Hipparcos} measurements of the star: the period $P$, time of mid-transit $Tc$, transit duration $w$ and the depth of the transit $d$. Bottom: ITP function for the first year after \textit{Hipparcos} observations, compared with the known transit light curve (orbital parameters derived using \citet{2000ApJ...532L..51C}). The significant peaks of the ITP fit mid-transit time; therefore, a single follow-up observation could have detected the transit.}
% \label{fig.HD209458hist}
% \end{center}
% \end{figure}

\label{lastpage}

\end{document}